\documentclass[a4paper,11pt]{article}
\usepackage{jheppub} %

\usepackage{package-notes}

\usepackage{subcaption}
\usepackage{bbold}

\newcommand{\im}{\mathrm{Im}}
\newcommand{\re}{\mathrm{Re}}

\usepackage[
]{} 

\title{The S-matrix bootstrap with neural optimizers I: \\
zero double discontinuity}

\preprint{LAPTh-061/24; CERN-TH-2024-209}

\author[a]{Mehmet Asım Gümüş,}
\author[a]{Damien Leflot,}
\author[a]{Piotr Tourkine,}
\author[b]{Alexander Zhiboedov}

\affiliation[a]{LAPTh, CNRS et Universit\'{e} Savoie Mont-Blanc 
9 Chemin de Bellevue, F-74941 Annecy, France}
\affiliation[b]{CERN, Theoretical Physics Department, CH-1211 Geneva 23, Switzerland}

\abstract{%
In this work, we develop machine learning techniques to study nonperturbative scattering amplitudes. We focus on the two-to-two scattering amplitude of identical scalar particles, setting the double discontinuity to zero as a simplifying assumption. Neural networks provide an efficient parameterization for scattering amplitudes, offering a flexible toolkit to describe their fine nonperturbative structure.
Combined with the bootstrap approach based on the dispersive representation of the amplitude and machine learning's gradient descent algorithms, they offer a new method to explore the space of consistent S-matrices. We derive bounds on the values of the first two low-energy Taylor coefficients of the amplitude and characterize the resulting amplitudes that populate the allowed region.
Crucially, we parallel our neural network analysis with the standard S-matrix bootstrap, both primal and dual, and observe perfect agreement across all approaches.
}

\begin{document}
\maketitle

\section{Introduction}
Machine learning techniques have experienced remarkable success in solving a wide range of complex tasks in recent years~\cite{lecun2015deep,schmidhuber2015deep,Goodfellow-et-al-2016}. A few notable examples include image recognition~\cite{krizhevsky2012imagenet}, mastering the game of Go~\cite{silver2016mastering}, and natural language processing \cite{brown2020language}. There are also numerous applications of machine learning in high-energy physics~\cite{Feickert:2021ajf}, and in string theory~\cite{Ruehle:2020jrk}. 
In this paper, we apply non-linear optimization techniques relying on machine learning to significantly improve an approach to the S-matrix bootstrap,  first introduced and developed by Mandelstam and Atkinson~\cite{Mandelstam:1958xc,Mandelstam:1959bc,Atkinson:1968hza,Atkinson:1968exe,Atkinson:1969eh,Atkinson:1970pe,Atkinson:1970zza}, that we introduce next.

The modern S-matrix bootstrap program~\cite{Kruczenski:2022lot} aims at charting the space of S-matrices consistent with analyticity, crossing symmetry, and unitarity, (\textbf{ACU}).
The standard modern approach~\cite{Paulos:2016fap,Paulos:2016but,Paulos:2017fhb,He:2021eqn,Chen:2022nym,EliasMiro:2022xaa} hinges upon convex optimization to explore the space of low-energy observables consistent with \textbf{ACU}.\footnote{See, however, \cite{Guerrieri:2024jkn} for a recent use of non-linear optimization to combine the S-matrix bootstrap with QCD experimental data.}
The Atkinson-Mandelstam approach proceeds differently and proposes to use multi-particle processes, ``inelasticity'', as an input, that needs to be modeled, to output a two-to-two scattering amplitude with this given inelasticity.
It proceeds by solving a set of \textit{non-linear} unitarity equations, written in a crossing-symmetric form that respects analyticity through the Mandelstam representation.
Due to the {non-linear nature} of these equations, convex optimization is {not obviously applicable}, and the strategy initially proposed by Mandelstam was an iterative fixed-point method -- the most basic tool of nonlinear functional analysis. 

In previous works~\cite{Tourkine:2021fqh,Tourkine:2023xtu}, the Atkinson-Mandelstam approach was implemented numerically in $d=2$ and $d=4$ spacetime dimensions, and two iterative methods for solving the non-linear equations were used: fixed-point iterations (in $d=2,4$) and Newton's method (in $d=2$). An important conclusion of these works is that the convergence range of these iterative strategies was, systematically, strictly narrower than the full allowed space of scattering amplitudes: the fixed points  were found to be attractive only in a small region of the parameter space, and repulsive everywhere else.

\begin{figure}
    \centering
    \includegraphics[]{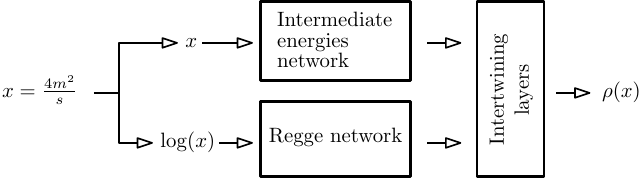}
    \caption{Architecture of a neural network that we use. The output, $\rho(x)$, is the discontinuity of the amplitude. Two sub-networks allow us to describe the high and intermediate energies in finer detail.}
    \label{fig:cartoon-NN}
\end{figure}

In the present article, we introduce a new, \textit{non-iterative} strategy to the Atkinson-Mandelstam approach based on a \textit{neural optimizer}. 
By \textit{neural optimizer}, we {mean} a neural network (NN) whose parameters are adjusted via an adaptive gradient descent without relying on any external training data (i.e. ``unsupervised training'').
This method (i)~solves the convergence range problem and, as a bonus, (ii) gives a new, practical way to scan over the space of amplitudes.
In practice, we use the NN to parameterize the discontinuity of the amplitude, which gives us access to the full amplitude via dispersion relations. Schematically, we write the Atkinson-Mandelstam equations as $\rm{LHS}=\rm{RHS}$, and we use the gradient descent to adjust the parameters of the network to minimize $(\text{LHS}-\text{RHS})^2$ summed over the grid points.
This idea is entirely general, and it underlies physics-informed machine learning \cite{karniadakis2021physics}.
Our framework and code build upon and substantially extend a prior study that utilized neural networks for the S-matrix bootstrap~\cite{Dersy:2023job}.\footnote{For other S-matrix related works using machine-learning, see \cite{Mizera:2023bsw,Niarchos:2024onf,Bhat:2024agd}.} 
In particular, we developed a network architecture adapted to our problem, that has sub-networks dedicated to solving the problem in different regions: the bulk of energies, and far UV, or the Regge region, as depicted in the schematic form in Figure~\ref{fig:cartoon-NN}.
Our algorithm uses the machine learning library PyTorch~\cite{Paszke:2019xhz} and the adaptive gradient-descent solver Adam \cite{kingma2017adammethodstochasticoptimization}.
We attach our code to the arXiv submission in auxiliary files. 

For this initial study, we consider a simplified version of the Atkinson-Mandelstam problem in $d=4$: the  ``toy model'' introduced in \cite{Tourkine:2023xtu}, which consists of amplitudes where we set double discontinuity to zero.\footnote{Let us mention that we also implemented the neural optimizer in $d=2$: this extended the convergence range of the algorithm to the full space of amplitudes in $d=2$ and allowed to recover the analytically known solutions with one CDD factor~\cite{Paulos:2016but}.}
This yields an important simplification of the structure of the equations, retaining their non-linearity, thus providing a numerically lighter problem that is ideal as a test ground for our method. Even in this simplified model, the neural optimizer outperforms the iterative strategies considerably.

Physically, we expect these amplitudes to be approximations to theories where the S-wave dominates the scattering.  Interestingly, and despite the simplification, this problem turns out to be rather rich, and we observe the dynamical emergence of resonances, familiar by now in the nonperturbative S-matrix bootstrap studies~\cite{Oller:2020guq,Guerrieri:2023qbg,Guerrieri:2021ivu,Guerrieri:2022sod,Acanfora:2023axz,Gumus:2023xbs,EliasMiro:2022xaa,Kruczenski:2022lot,He:2024nwd,He:2023lyy}.

To assert the reliability of our numerical method, we systematically check our results with standard nonperturbative bootstrap methods~\cite{Paulos:2016fap,Paulos:2016but,Paulos:2017fhb,He:2021eqn,Chen:2022nym,EliasMiro:2022xaa}, which constitute another, essential part of this work.

\begin{figure}
    \centering
    \begin{subfigure}[t]{0.4\linewidth}
    \hspace{-2.2cm}
    \includegraphics[scale=0.9]{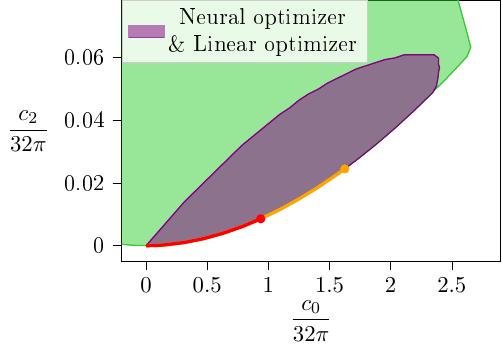}
    \caption{The single-disc almond.}
    \label{fig:results}
\end{subfigure}
\hspace{1cm}
\begin{subfigure}[t]{0.4\linewidth}
\hspace{-0.8 cm}
    \includegraphics[scale=0.9]{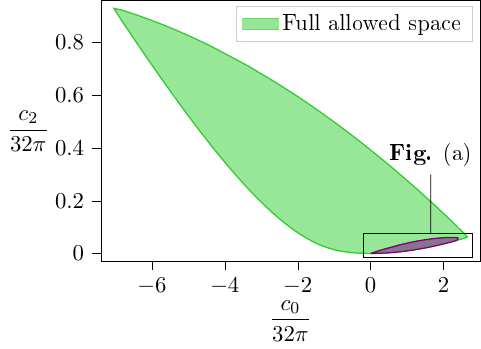}
    \caption{The full almond.}
    \label{fig:full_region}
\end{subfigure}
\caption{Plots of allowed S-matrices in the space of their Taylor coefficients $c_0$ and $c_2$ defined in \eqref{eq:c0c2def}.
\underline{Left.} {\color{purple}Purple region}: space of allowed S-matrices {with zero double discontinuity} {studied in this paper}: ``the single-disc almond''. 
All four methods (neural optimizer/standard bootstrap; primal/dual) produce identical results; hence, we show a unique shape. 
The {\color{red} red} and {\color{orange} orange} lines represent, respectively, the fixed-point iteration and Newton's method convergence range along the lower boundary, illustrating the superiority of the gradient-descent neural optimizer.
\underline{Right.} Comparison with the full space of S-matrices as worked out in~\cite{Chen:2022nym,EliasMiro:2022xaa}, or, equivalently, ``the full almond''. Our results shed new light on the role of the double discontinuity for the amplitudes obtained in these papers. In particular, in the complement of our region, surprisingly, the double discontinuity is required to unitarize even the S-wave. Characterizing this double discontinuity is an open problem.}
\label{fig:mainres}
\end{figure}

\paragraph{Main results.}
The main results of our paper are two-fold. 
Firstly, we demonstrate how a neural optimizer can successfully bypass two difficulties of the Atkinson-Mandelstam program: divergence of the iterative strategies and scanning the parameter space of theories.
Our method provides a new, flexible and promising tool in the modern S-matrix bootstrap machinery. 
Secondly, we characterize the space of scattering amplitudes for external scalars with no double discontinuity. 
{Our results follow from four numerical strategies: the neural optimizer and standard nonperturbative bootstrap, and for each we use the two standard optimization strategies: primal and dual.}
The analysis is summarized in Figure~\ref{fig:mainres}, 
in which we represent the amplitudes by the first few coefficients of its Taylor expansion around the crossing-symmetric point $s_0=t_0=u_0=\frac{4 m^2}{3}$
\begin{equation}
\label{eq:c0c2def}
\begin{aligned}
    T(s,t) \underset{s,t,u\to s_0}{=} c_0 + \frac{c_2}{m^4}\Big((s-s_0)^2+(t-t_0)^2+(u-u_0)^2\Big) + \dots\;.
\end{aligned}
\end{equation}
which is equivalent to the definition
\begin{equation}
c_0\equiv T(s_0,t_0),\quad c_2 \equiv \dfrac{m^4}{4}\partial_s^2 T(s,t)\bigg|_{s=t=s_0} \ .
\end{equation}
This study also teaches some lessons about the role of the double discontinuity of the amplitude, whose complete characterization is still an open question today, which we discuss in the discussion section.

\paragraph{Plan of the paper.}
In Section~\ref{sec:toy-model}, we provide an overview of scattering amplitude fundamentals and present our toy model. In Section~\ref{sec:nonlinear_opt}, we introduce the neural optimization framework and discuss its properties. In Sections~\ref{sec:Dual} and~\ref{sec:Primal}, we present our results for dual and primal strategies and compare them with the standard bootstrap techniques. In Section~\ref{sec:Discussion} we recap our results and discuss some physics implications. In Section~\ref{sec:numerical}, we discuss the numerical strategies presented in the paper. In Section~\ref{sec:FutureDirections}, we discuss the future directions of this program. Lastly, in the appendices, we give technical details on the neural network and review the standard nonperturbative bootstrap and semi-definite programming. 

\section{Amplitudes with zero double discontinuity}
\label{sec:toy-model}

In this paper, we study two-to-two scattering of identical scalars of mass~$m$. The variables $s,t,u$ are the usual Mandelstam variables and we have $s+t+u=4m^2$.
We further assume that there is no cubic self-coupling and no bound states below the two-particle threshold: any non-trivial properties of the amplitudes, such as resonances, will be generated ``dynamically'' by the algorithm.

Our ansatz for the amplitude is given by the following crossing-symmetric dispersive representation
\begin{equation}
T(s,t)=c_0+\frac1\pi\int_{4m^2}^{\infty}\!\!\!\!\mathrm{d}s'\dfrac{\rho(s')}{s'-s_0}\left(\dfrac{s-s_0}{s'-s}+\dfrac{t-t_0}{s'-t}+\dfrac{u-u_0}{s'-u}\right)\,,
\label{eq:def_toy_model}
\end{equation}
where the integral starts at the two-particle threshold $s=4m^2$, where the amplitude develops an imaginary part.
The extra denominator term ${s'-s_0}$ is called a subtraction term and forces that the value of the amplitude at the crossing-symmetric point $s=t=u=s_0$ is given by $c_0$
\begin{equation}
\label{eq:subtraction}
    T(s_0,t_0)=c_0
\end{equation}

It is immediate to see that the discontinuity of the amplitude in the $s$-channel for instance is given by the spectral function $\rho(s)$
\begin{equation}
{\rm Disc}_s T(s,t) \equiv \frac{T(s+i \epsilon) - T(s-i \epsilon)}{2 i} = \rho(s) .
\end{equation}
and likewise by $\rho(t)$ and $\rho(u)$ in the other channels.

Since the $s$-channel discontinuity does not depend on $t$, one can not take an extra discontinuity in another channel, therefore, in this ansatz the \textit{double} discontinuity of the amplitude vanishes:
\begin{equation}
\rho(s,t)\equiv {\rm Disc}_t {\rm Disc}_s T(s,t)  = 0 \ . 
\end{equation}
The double discontinuity of the amplitude is essential to describe scattering at fixed impact parameters and to unitarize partial waves with $J>0$, see \cite{Correia:2020xtr}. 

The Mandelstam representation~\cite{Mandelstam:1958xc} is a representation similar to~\eqref{eq:def_toy_model} that includes double-dispersive integrals of the double-discontinuity, see e.g.~\cite{Correia:2020xtr}. Our ansatz is, therefore, nothing but the actual Mandelstam representation, where the double discontinuity has been set to zero. We discuss the full representation in the discussion; see Eq.~\eqref{eq:mandelstamRep1}.

In the following parts of this section, we describe some properties of the amplitudes described by our ansatz. This information will be relevant for defining the allowed space, the single-disc almond of Figure~\ref{fig:results}, and for parametrizing the amplitudes.

\subsection{Unitarity}
To discuss unitarity we introduce the partial wave expansion of the amplitude
\begin{equation}
\label{eq:partialwaveexp}
T(s,t) =16 \pi \sum_{J=0}^\infty (2J+1) f_J(s) P_J(\cos \theta), ~~~ \cos \theta = 1 + \frac{2 t}{s - 4 m^2} \ , 
\end{equation}
where $P_J(\cos \theta)$ are the standard Legendre polynomials and $\cos \theta$ is the scattering angle. 
It is convenient to define the full partial wave as follows
\begin{equation}
    S_J(s)=1 + i \,\varphi(s) \, f_J(s)
    \label{eq:partial_wave}
\end{equation}
where $1$ comes from the free part of the S-matrix, and the phase-space factor $\varphi(s)$ is defined as
\begin{equation}
\varphi(s) \equiv \dfrac{\sqrt{s-4m^2}}{\sqrt{s}} \ .
    \label{eq:phase_space}
\end{equation}
in $d=4$. In this basis, unitarity is diagonalized and reads
\begin{equation}
 |S_J(s)| \leq 1 , ~~~ s \geq 4 m^2\,,    
\end{equation}
making transparent the statement that the probability of scattering in the spin-$J$ partial wave cannot exceed~$1$.
In the case where only two-particle states are available below some multi-particle scale $s_{\rm{MP}}$, unitarity is saturated
\begin{equation}
\label{eq:elastic-unitarity}
    |S_J(s)|=1\,,{\quad s<s_{\rm{MP}}}\,.
\end{equation} 
This condition is called \emph{elastic unitarity}. 

Let us now specialize to our ansatz~\eqref{eq:def_toy_model}. Taking the imaginary part of \eqref{eq:partialwaveexp} and using that $\im T(s,t) = \rho(s)$ does not depend on $t$,  
we conclude that the only way for the imaginary part of \eqref{eq:partialwaveexp} to be $t$-independent is that 
\begin{equation}
    \im f_{J>0} (s) = 0 .
\end{equation}
Just looking at the definition~\eqref{eq:partial_wave}, this immediately implies a violation of unitarity in~$J>0$ partial waves 
\begin{equation}
    \label{eq:non-unit}
    \rho(s,t) = 0 \implies |S_J(s)| \geq 1 \quad\textrm{for~} J>0\,.
\end{equation}

Within our model, only unitarity of the $J=0$ partial wave~$S_0(s)$ can be imposed, which in terms of the $f_0(s)$ reads:
\begin{equation}
    |S_0(s)|\leq 1\quad \Leftrightarrow \quad 2 \, \im f_0(s) \geq \varphi(s)\left(\im f_0(s)^2+\re f_0(s)^2\right),\quad s \geq 4m^2.
    \label{eq:S0_uni}
\end{equation}

By using the dispersive integral above \eqref{eq:def_toy_model}, we can express $\re f_0(s)$ for $s\geq4m^2$ as a function of $c_0$ and $\im f_0(s)$ as follows
\begin{align}
        \im f_0(s)= &{
        \frac{1}{16\pi}} \rho(s)\label{eq:imF0-rho} \ , \\
        \re f_0(s) =&\dfrac{c_0}{16\pi}+\mathrm{P.V.}\!\!\!\int_{4m^2}^{\infty}\!\!\dfrac{\mathrm{d}s'}{\pi}\dfrac{\im f_0(s')}{s'-\frac{4m^2}{3}}\left(\dfrac{s-\frac{4m^2}{3}}{s'-s}+K_0^{(4)}(s',s)\right),\label{eq:projected_disp}
\end{align}
where $K_0^{(4)}(s',s)$ is a kinematical kernel that reads\footnote{We kept the upper index $d=4$ to match the conventions of~\cite{Correia:2020xtr}.}
\begin{equation}
\label{eq:k0-def}
    K_0^{(4)}(s',s) = 2\left(\dfrac{s'-\frac{4m^2}{3}}{s-4m^2}\log\left(\dfrac{s+s'-4m^2}{s'}\right)-1\right).
\end{equation}
Eq.~\eqref{eq:imF0-rho} shows that one can interchangeably think of $\im f_0(s)$ or $\rho(s)$ as the function which parametrizes the full amplitude. Eq.~\eqref{eq:projected_disp} finally shows that the unitarity of $S_0$ can be written as a non-linear condition on a single function $\im f_0(s)$. For elastic unitarity, this inequality becomes the following equality for $s\geq4m^2$
\begin{equation}
    \label{eq:full-unit-rho}
    \im f_0(s) =\frac{\varphi(s)}{2}
    \left[
    \im f_0(s)^2+ 
        \left(\dfrac{c_0}{16\pi} + \mathrm{P.V.}\!\!\!\int_{4m^2}^{\infty}\!\!\dfrac{\mathrm{d}s'}{\pi}\dfrac{\im f_0(s')}{s'-\frac{4m^2}{3}}\left(\dfrac{s-\frac{4m^2}{3}}{s'-s}+K_0^{(4)}(s',s)\right)
        \right)^2
    \right]\;.
\end{equation}
In this paper, we use the neural optimizer to solve this non-linear functional equation. In the spirit of the Atkinson-Mandelstam approach, an inelasticity profile could be added to the RHS, and, in this case, we could solve for $\im f_0(s)$, given that profile~\cite{Tourkine:2023xtu}.

\subsection{The threshold behavior}
The behavior near the two-particle threshold $s\to 4m^2$ is constrained by unitarity \eqref{eq:S0_uni}.
Two possible behaviors are allowed, following from the fact that $\varphi(s) \sim \sqrt{s-4m^2}$ close to the threshold. It is either regular
\begin{subequations}
\label{eq:AsymptoticThreshold}
\begin{equation}
    \im f_0(s)\underset{s\to 4m^2}{\sim}\sqrt{s-4m^2},
\label{eq:AsymptoticThreshold-reg}
\end{equation}
or singular
\begin{equation}
    \im f_0(s)\underset{s\to 4m^2}{\sim}
        \dfrac{1}{\sqrt{s-4m^2}}.
\label{eq:AsymptoticThreshold-sing}
\end{equation}
\end{subequations}
The coefficient {of the singularity} in the right-hand side of \eqref{eq:AsymptoticThreshold-sing}  is fixed to $4m$ if elastic unitarity is imposed {near the two-particle threshold.}

\subsection{The Regge limit}
Let us first notice that unitarity \eqref{eq:S0_uni} immediately implies the following upper bound on $f_0(s)$
\begin{equation}
    |f_0(s)|\leq
    \frac{2}{\varphi(s)}
    ,\quad s>4m^2,
\end{equation}
and at high energies, we get
\begin{equation}
\label{eq:HEunit}
    \lim_{s\to\infty}|f_0(s)|\leq 2.
\end{equation}
A priori, the subtraction and~\eqref{eq:HEunit} allow for $\rho(s)$ to go to a constant at infinity. Here, we argue briefly that unitarity together with the dispersive presentation of the partial wave \eqref{eq:projected_disp} actually imply that
\begin{equation}
\begin{aligned}
    \lim_{s\to\infty}\im f_0(s) &= 0,\\
    \lim_{s\to\infty}\re f_0(s) &= 0.
    \label{eq:ReggeLimits}
\end{aligned}
\end{equation}
Imagine that the first equation is not true and $\im f_0(s) \sim \rho(s) \to \text{const}$ at high energies. 
By plugging this asymptotic behavior in the dispersive integral we get at high-energy
\begin{equation}
\label{eq:ref0-logs}
    \re f_0(s) \sim \int_{s_0}^{\infty}\!\!\dfrac{\mathrm{d}s'}{\pi}\dfrac{1}{s'-\frac{4m^2}{3}}\left(\dfrac{s-\frac{4m^2}{3}}{s'-s}+K_0^{(4)}(s',s)\right)\sim \log s \ ,
\end{equation}
which  violates \eqref{eq:HEunit} logarithmically. The same argument rules out any $\rho(s)$ which grows at high energies.
\footnote{\label{fn:unsubtractions}
The authors of~\cite{Chen:2022nym,EliasMiro:2022xaa} consider a space of amplitudes with two subtractions. For us, by the argument we just explained, an extra subtraction would not enlarge the space of allowed functions as $\rho(s)$ has to vanish at infinity anyway. A possible linear growth of the \textit{amplitude} that the second subtraction would a priori allow is actually impossible. The extra  term it would yield in Eq.~\eqref{eq:def_toy_model} has to be proportional to $s+t+u=4m^2$ via crossing symmetry and hence could be removed by changing the definition of $c_0$. Therefore, the space of amplitudes described in this paper has to be identical to the space of twice-subtracted amplitudes with no double discontinuity.}

Given $\im f_0(\infty) =0$, {equation~\eqref{eq:S0_uni} immediately implies $\re f_0(\infty)=0$, and therefore that $S_0(s)\xrightarrow[s\to\infty]{}1$.}\footnote{This statement follows from unitarity alone, elastic unitarity is not needed.} This proves Eq.~\eqref{eq:ReggeLimits}. 
The condition $\re f_0(\infty)=0$ in \eqref{eq:projected_disp} leads to the following sum rule
\begin{equation}
   \dfrac{c_0}{16\pi}-\dfrac{3}{\pi}\!\!\int_{4m^2}^{\infty}\!\!\!\!\mathrm{d}s'\,\dfrac{\im f_0(s')}{s'-\frac{4m^2}{3}}=0 \,.
   \label{eq:c0-sumrule}
\end{equation}
that fixes $c_0$ in terms of a dispersive integral of $\im f_0(s)$. It readily implies that
\begin{equation}
    \label{eq:c0-pos}
    c_0 \geq 0 \ ,
\end{equation} 
where $c_0=0$ corresponds to $T(s,t)=0$. 

Let us review next the leading correction to the asymptotic high-energy value of the amplitude, as worked out in~\cite{Tourkine:2023xtu}. It was found that if we impose that the scattering at high energies is elastic, i.e. that \eqref{eq:S0_uni} is saturated, then we get that the leading large $s$ behavior
is uniquely fixed to be
\begin{equation}
    \im f_0(s) = \dfrac{2\pi^2}{9\log(s)^2}+O\left(\tfrac1{\log(s)^3}\right),\label{eq:AsymptoticRegge}
\end{equation}
up to higher order logarithmic corrections, also worked out in in~\cite{Tourkine:2023xtu} up to a few orders. Notice that this is the slowest decay compatible with the sum rule \eqref{eq:c0-sumrule},
since the integral $\int^\infty ds \, 1/(s \log(s)^p)$ converges as long as $p>1$. If we relax elastic unitarity and allow for some inelasticity at high energies, faster decaying solutions are admissible. 
A power-law decay is for instance allowed, $\im f_0(s) \sim \frac{1}{s^{\alpha}}$, which would produce inelastic S-matrices $|S_0(s)|<1$. However, asymptotically, they still have to obey $S_0(s)\xrightarrow[s\to\infty]{}1$.

{Finally, let us also write down the dispersive representation of $c_2$. Using its definition in \eqref{eq:c0c2def} and the ansatz \eqref{eq:def_toy_model} we get the following sum rule}
\begin{equation}
        c_2=16\!\int_{4m^2}^\infty\!\!\!\!\mathrm{d}s'\dfrac{\im f_0(s')}{(s'-\frac{4m^2}{3})^3}\;.\label{eq:c2-sumrule}
\end{equation}
which in particular implies
\begin{equation}
c_2\geq 0 \,, 
\label{eq:c2-positive}
\end{equation}
where $c_2=0$ corresponds to $T(s,t)=0$.

Having described the essential properties of our amplitudes, we now describe the machine learning setup.

\section{Neural optimizer}
\label{sec:nonlinear_opt}

In this section, we review a few elements related to optimization and machine learning, and explain our setup.

Linear and semi-definite optimization are convex optimization problems: they are supported by a robust theoretical framework, with established algorithms and theorems (e.g. simplex, interior point), and provide efficient and reliable solutions~{\cite{Boyd:2004fnq}}. These methods are frequently used in the bootstrap community, for instance, through the use of the software SDPB~\cite{Simmons-Duffin:2015qma}.
However, the basic feature of the Atkinson-Mandelstam approach is its non-linearity.

This immediately yields a numerical challenge since non-linear optimization is a much more complex terrain than convex optimization. It can be highly sensitive to initial conditions, may be affected by multiple local optima, and generally requires a blend of heuristic strategies, numerical techniques, and intuition to navigate efficiently. 

Machine learning algorithms based on neural networks have proven powerful tools for performing such tasks. They combine efficient gradient-descent algorithms with the neural networks' flexibility.

A few principles back the efficiency of these machine-learning methods. First of all, neural networks are known to be ``universal approximators'' \cite{cybenko1989approximation,hornik1989multilayer}, and thus can in principle fit any function.
Neural networks are also remarkably efficient at finding local minima that are often close to the true global minimum. At first glance, this might sound paradoxical why given the enormous number of parameters in NNs. Intuitively, one might expect such over-parameterization to increase the risk of overfitting. However, the opposite is observed. This phenomenon can be understood in the infinite width limit, where NNs can be mapped onto a well-studied algorithm known as a ``kernel machine'' \cite{Jacot:2018ivh,lee2017deep} (see also the earlier work in~\cite{Neal1996} and the extension to infinite depth in~\cite{matthews2018gaussian}). In this limit, as the dimensionality becomes infinite, gradient descent effectively corresponds to moving within a bowl-shaped potential. This insight aligns with the intuitive notion that, in higher-dimensional spaces, the local minima encountered by NNs tend to become shallower, reducing the risk of poor optimization outcomes.
A perturbation theory in a variable $\frac{\rm{depth}}{\rm{width}}$ {was later developed in}~\cite{roberts2022principles}. To our understanding, whether these results extend far from the infinite width and large depth limits (as is the case for us) is an open question.

Another important aspect of machine learning libraries is that they are easy to use, {open source}, well-maintained, and highly optimized. For instance, they all feature \textit{automatic differentiation}, a tool that computes the gradient at the same time as the loss function using the \textit{backpropagation} algorithm, which we explain below, and that yields a significant improvement in computational speed.

Before describing our precise architecture, let us close these introductory remarks with a few bibliographical comments. 
Besides the references focused on machine-learning mentioned in the introduction, let us also mention that other non-linear techniques are being developed for the S-matrix bootstrap. For instance, in the work~\cite{Guerrieri:2024jkn}, a gradient-free algorithm, known as Particle Swarm Optimization, was used alongside standard S-matrix bootstrap techniques to model the pion-pion scattering amplitude. The non-linear optimization enables navigation of the non-convex $\chi^2$ landscape and fitting QCD data.

\subsection{Training and loss function}
\label{sec:training}

Let us introduce the basic terminology of machine learning. A neural network is a function, $x\mapsto \mathrm{NN}_{\sigma}(x)$, parametrized by a set of internal parameters collectively referred to as $\sigma$. These parameters, commonly known as \textit{weights} and \textit{biases}, can be thought of as abstract variables that characterize the network's behavior. Their precise definition will be provided following Eq.~\eqref{eq:lin-layer}.
During the \textit{training}, $\sigma$ are tuned to minimize a quantity called the \textit{loss function} of the network, or simply the loss: $\mathcal{L}(\mathrm{NN}_\sigma)$.

The choice of the loss function $\mathcal L$ is determined by the problem we are solving. 
In the dual bootstrap approach, described in Section~\ref{sec:Dual}, this quantity is given by the so-called dual Lagrangian. For the primal bootstrap problem, this quantity is directly given by the unitarity equation~\eqref{eq:elastic-unitarity} {written as $\mathcal L \sim (\text{LHS}-\text{RHS})^2$}. Importantly, in both cases, no input data is required, as the loss is computed exclusively from the neural network itself. Therefore, we are within the realm of \textit{unsupervised} training.

The training phase consists of a succession of \textit{epochs}.
An epoch represents a full iteration during which the network's parameters are updated using a step of gradient descent. Each epoch begins with the evaluation of the loss function $\mathcal{L}$, and a simultaneous evaluation of its gradient 
\begin{equation}
    \nabla_\sigma\mathcal{L},
    \label{eq:backpropagation}
\end{equation}
This gradient is then passed to a gradient-based optimizer, which updates the parameters $\sigma$ to minimize the loss. The process repeats with a new evaluation of the loss in the next epoch. It is crucial to note that the gradient is, in general, evaluated \textit{at no extra computational cost}, thanks to the backpropagation algorithm. This algorithm is used to obtain the gradient of the neural network function. It is a standard machinery based on a simple application of the chain rule. Let us explain in a few words how it works.
In general, the neural network is a composition of many elementary functions, $f_1 \circ f_2 \circ \dots\circ f_n(x)$. 
As the algorithm evaluates the output of the NN for a given $x$, it also computes the derivative of $\text{NN}_{\sigma}(x)$ with respect to $\sigma$ by using the chain rule, which multiplies these derivatives \textit{in reverse}, hence the name, backpropagation. This derivative is used to compute the gradient of the loss function and perform the gradient descent.
Training continues until the loss reaches a plateau, indicating that further epochs no longer lead to significant improvement. 

\subsection{Neural network architecture} 
\label{sec:architecture}
Let us now outline the general architecture of neural networks and the specific design of our model.
Neural networks are built from a succession of \textit{layers}. The standard architecture, which we also use, is an alternation of two kinds of layers:

\medskip

\noindent\underline{Linear layers.}
\begin{equation}
\label{eq:lin-layer}
    \vec{z}\mapsto W\cdot \vec{z}+\vec{B}.
\end{equation}
A linear layer takes the vector input $\vec{z}$ and multiplies it by the weights matrix $W$ and adds the biases vector $\vec{B}$. The matrix elements of $W$ and $\vec{B}$ are the internal parameters that are tuned during the training of the neural network. Note that the dimensionality of $\vec z$ and $ W\cdot \vec{z}+\vec{B}$ do not have to match, and they can be equal to one.

\medskip

\begin{figure}[h!]
    \centering
    \includegraphics[]{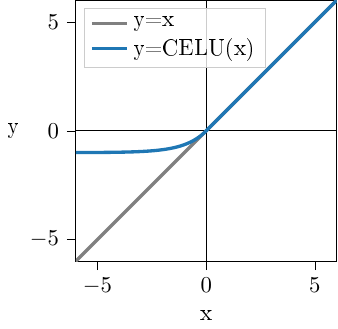}
    \caption{Plot of the activation function $\mathrm{CELU}$ defined in \eqref{eq:CELU_function}.} 
    \label{fig:CELU}
\end{figure}

\noindent\underline{Activation layers.}
\begin{equation}
\label{eq:activ-layer}
    \vec{z}\mapsto \vec C(\vec{z}).
\end{equation}
The activation layers apply a given activation function $\vec C$ element-wise to the input vector~$\vec{z}$. The default activation function is the so-called $\mathrm{ReLU}:x\mapsto\max(0,x)$. In this work, we used the $\mathrm{CELU}$ function, see Figure~\ref{fig:CELU}, which is a differentiable version of $\mathrm{ReLU}$:
\begin{equation}
    \mathrm{CELU}(x)\!=\!\max (0,x) + \min (0, \exp(x)-1).
    \label{eq:CELU_function}
\end{equation}
We observed that using this $\mathrm{CELU}$ activation function speeds up the training phase compared to the standard ReLU. 
One possible explanation for this might be linked to the fact that we expect the functions approximated by the neural network to be reasonably smooth and, therefore, the use of smooth building blocks {guides the gradient descent efficiently}.

Overall, after passing through two such layers, an input $\vec z$ is mapped to a new vector of potentially different dimensionality
\begin{equation}
    \label{eq:y-twolayers}
    \vec z\mapsto \vec C(W\cdot \vec z+\vec B)
\end{equation}
Each individual component of $\vec C(\dots)$ a priori depends on all entries in $\vec z$. 
Since these layers often appear in pairs, we will refer to the combination of a linear layer and an activation layer as a {layer block}:
\begin{quote}
    \centering
    \underline{layer block} = linear layer + activation layer
\end{quote}

The \textit{depth} of a neural network is determined by the number of linear layers; the layers between the first and the final one are called \textit{hidden} layers. The width of a layer is the size of its output.\footnote{Each hidden layer may have a different width.}

\begin{figure}
    \centering
    \includegraphics[]{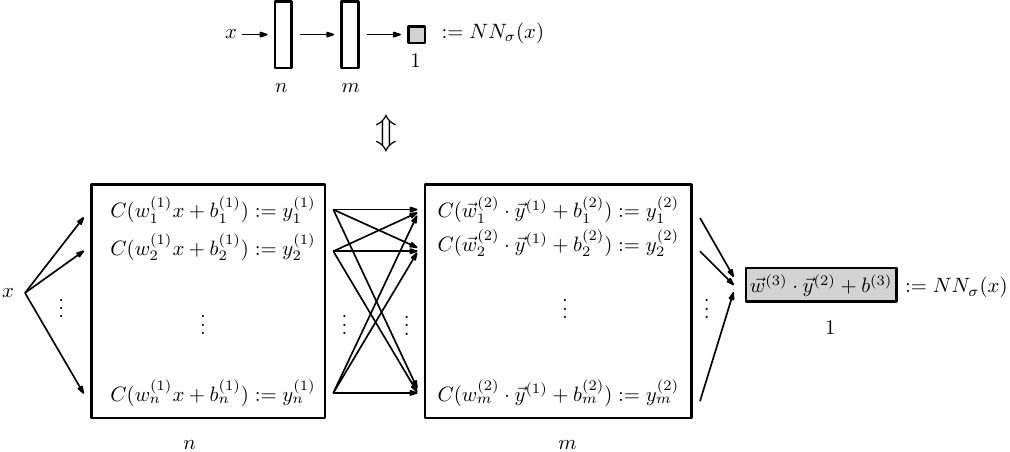}
    \caption{Example of a neural network with 2 layer-blocks with widths $n$, $m$ and the final layer (linear layer) returning a single output (width $1$). In these graphical conventions, the last layer is always of width $1$. The \underline{top} and \underline{bottom} picture represent {the same network}: the top picture is a condensed notation, which we use in Figure~\ref{fig:NN} to describe our full architecture, and the bottom is the ``definition'' of the condensed picture.
    The function $C$ is the activation function, and the $w$'s and $b$'s are the weights and biases. They have a superscript, which represents the layer, an index, which represents the row number, and arrows when they are vectorial quantities. These indices are given in this picture for the sake of definiteness. Lastly, in this work, $C(\cdot)\equiv\rm{CELU}(\cdot)$, defined in Eq.~\eqref{eq:CELU_function}.}
    \label{fig:NN-example}
\end{figure}

A typical network with two layer blocks and a final layer is represented in Figure~\ref{fig:NN-example}. In this example the final layer, in gray, consists of a linear layer only.
Note that for us, the initial input is typically a single point $x$, not a grid of points. The output is the function~${\rm NN}_\sigma(x)$.

\subsection{Specific architectures for our problem}
\label{sec:spec_arch}

In the context of our problem, we examine functions of $s$ for $s\geq 4m^2$. For numerical convenience, we compactify this range using the transformation
\begin{equation}
    s\mapsto x=\dfrac{4m^2}{s}\in [\,0,1]\,.
\end{equation}
Here, $x$ will serve as the input to our networks.

\begin{figure}
    \centering
    \includegraphics[]{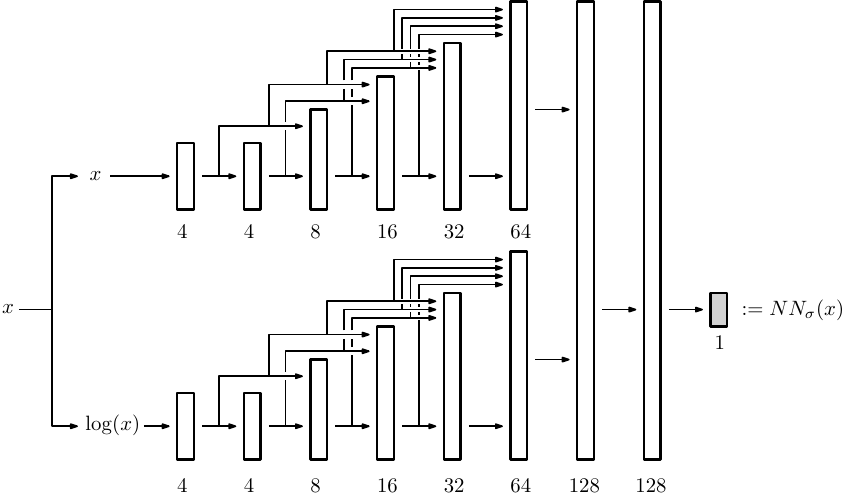}
    \caption{Full architecture of the neural network for the primal approach, with 8 layer-blocks, implementing skip connections. With the skip connections, layer blocks depend on several layers preceding them simultaneously (the altitude of the arrow in the box does not represent anything, all the variables from all relevant layers enter in a similar way). Intuitively each part of the network learns a specific physical regime: Regge/threshold.}
    \label{fig:NN}
\end{figure}

We employed two distinct neural network architectures to obtain the results presented in this paper.
The neural network we employ for the dual problem in the next section is of the same type as the one in Figure~\ref{fig:NN-example}. It consists of 6 layer-blocks, of constant width 64.
This simple architecture is sufficient for the dual setup, where the decaying behavior of the amplitude is already built-in, as explained in Section~\ref{sec:uv-ir-sens}. For the primal problem described in Section~\ref{sec:Primal}, we use a more complex architecture depicted in Figure~\ref{fig:NN} covering exponentially high energies. This architecture is essential to correctly capture the logarithmic decay, specifically $1/\log(s)^2$, via unitarization at high energies.

The primal bootstrap neural network takes two inputs: $x$ and $\log(x)$. These two inputs go through two independent sets of layer blocks or ``sub-networks''; the output of each independent sub-network is then concatenated and passes through two additional layer blocks before the final layer. Intuitively at high energy, when $x\ll 1$, only the output from the sub-network associated with~$\log(x)$ varies with~$x$, allowing them to learn the Regge behavior during training. Conversely, the low energy behavior is captured by the other sub-network associated with~$x$.\footnote{Curiously, a similar architecture is used by the NNPDF collaboration in the study of QCD parton distribution functions \cite{NNPDF:2021njg}.}

In this network, we also use \textit{skip connections}, which are connections that link a layer to several preceding layers.
Skip connections are commonly used to address the ``vanishing gradient problem''. This problem occurs in deep networks, where the gradient must pass through many layers via the chain rule, often becoming very small as it propagates. As a result, the deepest layers learn very little with each epoch. Skip connections allow the gradient to reach deep layers with fewer applications of the chain rule, helping to prevent the gradient from becoming too small. This enables the deeper layers to learn more {efficiently}. We observed that skip connections help the network converge in fewer epochs {and yield an almost two-fold speed-up.}

\subsection{Implementation, runtimes}
Initializing the NN's weights is important in achieving good convergence during training.\footnote{Biases are commonly initialized to zero.} If all weights are initialized with the same value, permutation symmetries will occur between weights within the same layer. Consequently, these weights receive identical gradients during each training epoch, causing them to remain equal throughout training. This effectively reduces the number of degrees of freedom in the NN, causing it to fail to learn complex behaviors. 
This is a known fact, and many solutions exist to mitigate this issue.
For this work, we used the Kaiming normal initialization~\cite{he2015delvingdeeprectifierssurpassing}, which is natively implemented in PyTorch. This method assigns random values for the weights, from normal distributions, breaking the permutation symmetry. The distributions are adjusted based on the widths of the layers to prevent the vanishing or exploding gradient problem.

The depths of our networks (6 and 9, respectively) were chosen to ensure that for any initialization, as described above, the loss reliably converges to a value of the same magnitude. 
To update the network's parameters at each epoch, we used the adaptive gradient descent algorithm Adam~\cite{kingma2017adammethodstochasticoptimization}, already implemented in PyTorch. The recommended values of $\beta_1$ and $\beta_2$ were used: $0.9$ and $0.999$, respectively. 
We only had to reduce the learning rate from its default value $10^{-3}$ to $10^{-4}$: at the default learning rate,\footnote{The learning rate is a quantity which controls how much the parameters of the NN are updated at each gradient step.} the loss exhibited stochastic jumps at each epoch with no clear convergence.\footnote{While the parameters $\beta_1$ and $\beta_2$ should, according to the authors of~\cite{kingma2017adammethodstochasticoptimization}, not be touched, adjusting the learning rate is done routinely.}

Lastly, we incorporated a learning rate \textit{scheduler} to improve training stability. Also provided by PyTorch, the scheduler gradually decreases the learning rate to $10^{-5}$ by the end of the training. Initially, a learning rate of $10^{-4}$ facilitates rapid loss reduction, while the smaller learning rate toward the end leads to finer tuning.\footnote{While these values work well for most inputs, smaller learning rates ($10^{-5}$ initially, reduced to $10^{-6}$) can sometimes yield better results, particularly for more sensitive setups.}

With these settings, training typically requires $10^5$ epochs which takes around 10 minutes on a standard laptop. 
Once training is completed, the weights and biases are saved in a separate file,\footnote{In PyTorch the network parameters are saved in a Python dictionary, and the file containing it is called a \textit{state dictionary.}} which allows for hot-starting the neural network in the future training runs. When the setup for a new training task is similar to the original, hot-starting significantly reduces the required number of epochs; typically by a factor of 2, compared to training the network from scratch.


\section{Dual optimization}
\label{sec:Dual}

The optimization problems associated with the S-matrix bootstrap fall into two main categories: primal and dual. In the next two sections, we discuss these strategies in detail.

In the \emph{dual} approach, the task is to demonstrate that certain values of the observables are \textit{incompatible} with \textbf{ACU}. As a simple example, consider the sum rules \eqref{eq:c0-sumrule} and \eqref{eq:c2-sumrule}. Combined with the unitarity condition  
$\im f_0(s)\geq 0$, the sum rules imply
\begin{equation}
    0\leq c_2\leq \frac{3}{64} c_0\,.
\label{eq:pos-wedge}
\end{equation}
Any pair of points $(c_0,c_2)$ that lies outside of the region defined by \eqref{eq:pos-wedge} is automatically \textit{ruled out}.

More generally, we can use the weak duality principle to derive the dual bounds~\cite{Boyd:2004fnq,Guerrieri:2021tak}. Consider the maximization problem of an objective function $\mathcal{P}(p_i)$ under a set of constraints on its variables $\{p_i\}$.
We sometimes refer to $\mathcal{P}$ as the \emph{primal objective}. 
We can express this problem by means of a Lagrangian $\mathcal{O}(p_i,d_i)$ consisting of $\mathcal{P}$ and the Lagrange multipliers~$\{d_i\}$ for each constraint. Then, the following max-min inequality holds:
\begin{equation}
\min_{\{d_i\}} \, \max_{\{p_i\}} \, \mathcal{O}(p_i , d_i) \quad \geq \quad \max_{\{p_i\}} \, \min_{\{d_i\}} \, \mathcal{O}(p_i , d_i) \, .
\label{eq:maxmin_ineq}
\end{equation}
Integrating out the variables $\{p_i\}$ on the LHS and staying on the support of vanishing constraints on the RHS, yield the \emph{dual objective} $\mathcal{D}$ and the maximum of our target respectively, while obeying the weak duality
\begin{equation}
\min_{\{d_i\}} \, \mathcal{D}(d_i) \quad \geq \quad  \max_{\{p_i\}} \, \mathcal{P} (p_i) \, .
\label{eq:weak_duality}
\end{equation}
Thus, values greater than the minimum of $\mathcal{D}$ are ruled out from being a solution to the maximization problem. The use of Lagrange multipliers in order to constrain the pion-pion scattering amplitude was first introduced in \cite{Lopez:1974cq} and later followed up in \cite{Lopez:1976zs,Bonnier:1975jz,Lopez:1975ca,Lopez:1975wf}. For a detailed discussion in the context of modern S-matrix bootstrap, see \cite{Guerrieri:2021tak}.

Let us now proceed to derive the dual objective $\mathcal{D}$ in our problem in order to chart the space of consistent S-matrices in the $(c_0,c_2)$ plane. Therefore, we seek which values of $(c_0,c_2)$ cannot correspond to S-matrices that satisfy \textbf{ACU}.\footnote{Remember that in our specific problem, \textbf{U} is only unitarity of the $J=0$ partial wave.}  
In the main text, for brevity, we explain how to bound the value of $c_0$, while keeping $c_2$ unrestricted. Generalizing it to the full $(c_0,c_2)$ plane is straightforward, and we provide the details in Appendix~\ref{sec:c0c2plane}.

\paragraph{Dual problem.} The first step is to write down the Lagrangian suited for our problem containing \textit{primal variables} $\{c_0,\re f_0(s),\im f_0(s)\}$ and Lagrange multipliers, which, from now on, we call as \textit{dual variables} $\{\kappa_0, w_0(s), \lambda_0(s)\}$. It reads:
\begin{align}
    \mathcal{O} &= c_0 + \kappa_0 \left[c_0 {-} \int^\infty_{4m^2} \!\!\! dv \, \frac{3}{\pi} \frac{ {n_0} \, \im f_0(v)}{(v{-}\tfrac{4m^2}{3})} \right] + \int^{\mu^2}_{4m^2} \!\!\! ds \, w_0(s) \, \mathcal{A}_0(s) + \int^\infty_{4m^2} \!\!\! dv \, \lambda_0(v) \, n_0^2 \, \det \mathcal{U}_0(v)
\label{eq:primal_lag}
\end{align}
where $$n_0 \equiv 16\pi\,,$$ is a normalization factor. The dual variables  in $\mathcal{O}$ enforce the \textbf{ACU} constraints, by ensuring the vanishing of the last three terms. Therefore, the maximum of $\mathcal{O}$ is, by definition, the maximum allowed value of $c_0$ for the problem. Let us examine each constraint in detail.

The first constraint is the sum rule~\eqref{eq:c0-sumrule}, which was a direct consequence of having unitarity in the Regge limit \eqref{eq:ReggeLimits}.
Analyticity is imposed in the second constraint by enforcing the dispersion relations between the real and imaginary parts of the partial waves
\begin{equation}
    \label{eq:roy_eq}
    \mathcal{A}_0(s) \equiv \re f_0(s) - \dfrac{c_0}{16\pi}-\mathrm{P.V.}\!\!\!\int_{4m^2}^{\infty} \!\!\! dv \, {n_0} \, k_{0,0}(s,v) \, \im f_0(v) \equiv 0 \, , \quad s \in [4m^2,\mu^2] \, .
\end{equation}
This is the defining equation for $\re f_0(s)$. Above in \eqref{eq:primal_lag}, we also introduced a finite cut-off on the analyticity constraints, which sets $\re f_0(s)$ to zero for $s>\mu^2$. In order to probe the unitarization in the UV, we will explore the dependence of the dual bounds on the value of $\mu^2$, and use $\kappa_0=0$ and $\kappa_0\neq0$ interchangably to quantify the effects of the $c_0$ sum rule. Eventually, we take the limit $\mu^2 \to \infty$.\footnote{In the full problem, by projecting a fixed-$t$ dispersion relation for the amplitude to higher partial waves, we get an analogous family of equations $\mathcal{A}_J(s) = 0$ called the \emph{Roy equations} \cite{Roy:1971tc,Ananthanarayan:2000ht}. If the amplitude has a non-vanishing double discontinuity, then the Roy equations have a limited validity range in $s$, and the maximally allowed cutoff is $\mu^2=60m^2$.}
For the normalization purposes, we defined the kernel ${n_0} \,  k_{0,0}(s,v) $ which encompasses the kernel \eqref{eq:k0-def} and the extra $1/(s'-s)$ and $1/(s'-s_0)$ parts of~\eqref{eq:projected_disp}. It reads
\begin{equation} {n_0} \, k_{0,0}(s,v)  = \frac{1}{\pi} \, \left( \frac{1}{v-s}-\frac{3}{v-4m^2/3}+\frac{2 \log \left(1+\frac{s-4m^2}{v}\right)}{s-4m^2} \right) .
\end{equation}
Finally, unitarity is imposed by demanding the following matrix to be positive semi-definite
\begin{equation}
    \label{eq:unitarity_matrix}
    \mathcal{U}_0(s) = 
    \begin{pmatrix}
        \varphi(s) \,\im f_0(s) \phantom{11} & \varphi(s) \, \re f_0(s) \\
        \varphi(s) \, \re f_0(s) \phantom{11} & 2 {-} \varphi(s) \, \im f_0(s) \ ,
    \end{pmatrix} \succeq 0 \ ,
\end{equation}
implying that $\det \mathcal{U}_0(s) \geq 0$, which is identical to the condition \eqref{eq:S0_uni}. Notice that semi-positivity in the unitarity condition implies $\lambda_0(s) \geq 0$. 

In the spirit of \eqref{eq:weak_duality}, we can start constructing the dual objective by integrating out~
$c_0$, and its equations of motion read
\begin{equation}
1 + \kappa_0 - \int_{4m^2}^{\mu^2} \!\!\! ds \, \frac{w_0(s)}{{ 16\pi }} = 0 \, .
\label{eq:Kappa_eqofmot}
\end{equation}
Similarly, the equations of motion for $f_0(v)$ give
\begin{equation}
\begin{aligned}
\varphi(v) \, \re f_0(v) &= \frac{1}{2 \lambda_0(v) n_0} \, (w_0(v)/n_0) \ , \\ 
\varphi(v) \, \im f_0(v) &= 1 + \frac{\overline{\mu}_0(v)}{2 \lambda_0(v) n_0} \ .
\label{eq:F0_eqofmot}
\end{aligned}
\end{equation}
They express the spin-zero partial wave in terms of the dual variables, and we defined the auxiliary function ${\overline \mu}_0(v)$ as follows
\begin{align}
{\overline \mu}_0(v) &\equiv -\frac{1}{\pi} \frac{3\kappa_0}{(v{-}\tfrac{4m^2}{3})} - \text{P.V.} \!\! \int^{\mu^2}_{4m^2} \!\!\! ds \, w_0(s) \, k_{0,0}(s,v) \ .
\label{eq:Mubar}
\end{align}
Substituting \eqref{eq:Kappa_eqofmot} and \eqref{eq:F0_eqofmot} in $\mathcal{O}$ leaves us with the dual objective
\begin{equation}
\mathcal{D} = \int^{\mu^2}_{4m^2} \frac{dv}{\varphi(v)} \frac{\left[w_0(v)/n_0\right]^2}{4 \lambda_0(v)} + \int^\infty_{4m^2} \frac{dv}{\varphi(v)} \frac{1}{4 \lambda_0(v)} \left[ 2 \lambda_0(v) n_0 + \overline{\mu}_0(v) \right]^2 \ .
\label{eq:duallagrangian}
\end{equation}

\paragraph{Optimal dual problem.} $\mathcal{D}$ is a positive functional over the set of functions $\{w_0(s),\lambda_0(s)\}$, and it evaluates to a strict upper bound on the objective $\mathcal{P}$ by virtue of~\eqref{eq:weak_duality}. Hence, we are interested in finding its minimum. To this aim, we can optimize once more, this time with respect to $\lambda_0$. Solving for its equation of motion on the positive branch gives: 
\begin{align}
\label{eq:opt_lambda0}
2 \lambda_0(v) n_0 &=
\begin{cases}
\sqrt{\overline{\mu}_0(v)^2 + (w_0(v)/n_0)^2} &\quad \text{if } v \leq \mu^2 \ , \\
\quad\quad |\overline{\mu}_0(v)| &\quad \text{if } v > \mu^2 \ .
\end{cases}
\end{align}
Plugging it back in \eqref{eq:duallagrangian} gives the optimal dual objective
\begin{equation}
\label{eq:dualbar}
\overline{\mathcal{D}} = \int^\infty_{4m^2} \frac{dv}{\varphi(v)} \, n_0 \left[ \, \overline{\mu}_0(v) + \sqrt{ \overline{\mu}_0(v)^2 + \left( w_0(v)/n_0 \right)^2 \theta\left(\mu^2-v\right) } \, \right]  \ ,
\end{equation}
given solely in terms of $w_0$, and where $\theta(v)$ is the Heaviside step function. An interesting outcome of optimality in $\lambda_0$ is that it leads to $|S_0(v)|=1$ for all $v \in [4m^2,\mu^2]$, which can be seen by plugging \eqref{eq:opt_lambda0} into \eqref{eq:F0_eqofmot}.
\setlength{\parindent}{20pt}

Notice that $\overline{\mathcal{D}}$ takes positive values (i.e. bounded from below) for any test function $w_0(s)$, and each time it returns a rigorous upper bound on the objective $\mathcal{P}$.
Moreover, it turns out to be convex, since the second order variation of $\overline{\mathcal{D}}$ with respect to $w_0$ is positive as well.
Hence, our task is now reduced to finding out suitable test functions on the support $[4m^2,\mu^2]$ that minimize $\overline{\mathcal{D}}$, i.e. to find
\begin{align}
\min_{\{ w_0 \}} \quad \overline{\mathcal{D}} \, .
\end{align}

However, $\overline{\mathcal{D}}$ is non-linear in $w_0$, since it contains a square root and an integral against $k_{0,0}$. It renders difficult the task of minimization over the space of $w_0$ via standard gradient descent algorithms.
This represents an ideal scenario in which the neural network can efficiently navigate a complex function space to identify the desired minimum.

In the next subsections, we minimize $\overline{\mathcal{D}}$ with the help of a neural network and compare it with a semi-definite programming solution.


\begin{figure}
    \centering
    \includegraphics[scale=1]{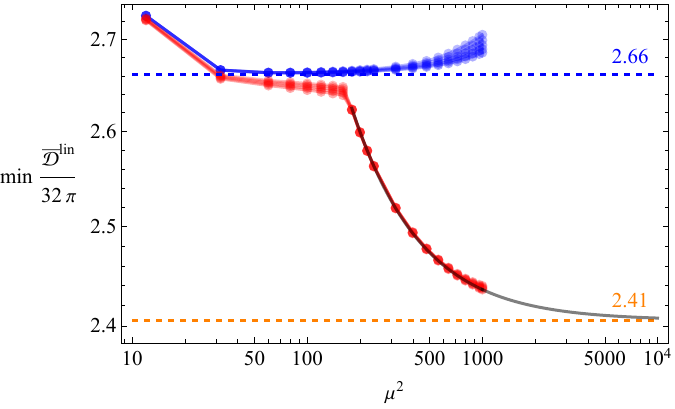}
    \caption{{Dual upper bound on $\max c_0/(32\pi)$ as a function of the finite analyticity cut-off $\mu^2$ in units of $m^2$.} {\color{blue}Blue}/{\color{red}red} data points show the bounds  {\color{blue}without ($\kappa_0=0$)}/{\color{red}with ($\kappa_0\neq 0$)} $c_0$ sum rule~\eqref{eq:c0-sumrule}, and increasing opacity values in each color stand for $N_\text{max}=80,\dots,96$. Extrapolation in $\mu^2$ (black line) agrees with the result of neural optimizer at $\mu^2 = \infty$ ({\color{orange}orange dashed}). Notice that the restriction from \eqref{eq:c0-sumrule} on the high energy behavior is crucial to reach convergence to the optimal bound after $\mu^2 \simeq 200m^2$.}
    \label{fig:Dual_extra}
\end{figure}

\begin{figure}
    \centering
    \includegraphics[scale=0.85]{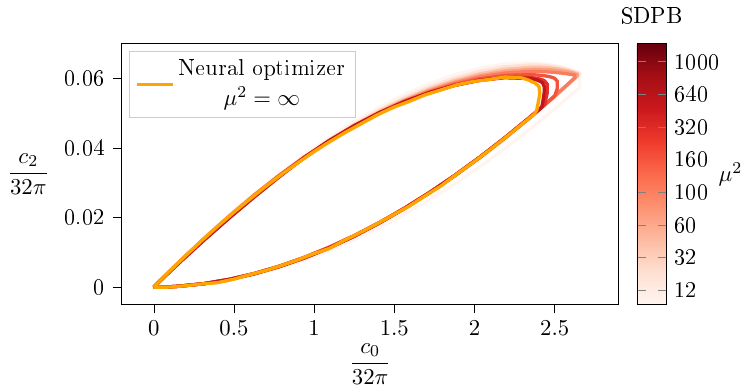}
    \caption{Two-dimensional dual exclusion regions obtained with SDPB (increasing opacities of {\color{red}red}) at finite $\mu^2$ and neural optimizer at $\mu^2 = \infty$ ({\color{orange}orange}).}
    \label{fig:Dual}
\end{figure}

\subsection{Neural optimizer}
\label{sec:Dual_NN}
As explained in Section~\ref{sec:nonlinear_opt}, during training, the neural network converges towards the function that minimizes the loss. This process is analogous to the dual optimization problem where $\overline{\mathcal{D}}$ must be minimized over the set of functions $w_0$. Thus, it is natural to parametrize the function $w_0$ using a neural network. The space of all possible $w_0$ functions is then restricted to the subset that can be represented by the neural network. For this problem, we use a neural network with 6 layer blocks, each of width 64, as explained in Section~\ref{sec:architecture}. We also set $\kappa_0\neq 0$ which imposes the sum rule~\eqref{eq:c0-sumrule}.

The loss is chosen such that $\overline{\mathcal{D}}$ is minimized during training: in our case, we use $\left(\overline{\mathcal{D}}\right)^2$ as the loss.\footnote{Any loss could work as long as it is minimized when $\overline{\mathcal{D}}$ is minimized.} 
We employed a piecewise-linear spline interpolation between grid points, allowing us to express the integrals as matrix multiplications, as originally done in~\cite{Paulos:2016but}. Details are provided in Appendix~\ref{sec:integration_matrix}.

The most straightforward way to apply the neural network to this problem is to define
\begin{equation}
    \begin{aligned}
    \label{eq:simpledualNN}
        w_0(v) &\equiv \mathrm{NN}_\sigma(4m^2/v),\quad v\geq4m^2\,,\\
        \mathcal{L}_D&\equiv \left(\,\overline{\mathcal{D}}\,\right)^2,
    \end{aligned}
\end{equation}
assuming $\overline{\mathcal{D}}>0$. With this setup, at the end of the training process, we obtain\footnote{The case $\min\overline{\mathcal{D}}<0$, as explained in Appendix~\ref{sec:c0c2plane}, can easily be treated by shifting $\overline{\mathcal{D}}$ by some sufficiently large positive value.}
\begin{equation}
    \min_{\sigma} \quad \mathcal{L}_D \,\geq\, \left(\min_{\{ w_0 \}} \quad \overline{\mathcal{D}}\right)^2 \, .\label{eq:dual_NN}
\end{equation}
The inequality in Eq.~\eqref{eq:dual_NN} arises from the fact the neural network we are using has a finite size (the set $\sigma$ is finite). Hence it cannot cover the entire space of function $\{w_0\}$.

The simple ansatz \eqref{eq:simpledualNN} can be improved to speed up training and enhance numerical stability by incorporating the expected Regge and the threshold behavior of $w_0(v)$. It is particularly relevant in the limit $\mu^2 \to \infty$ where it is important to ensure that the integral in Eq.~\eqref{eq:Kappa_eqofmot} remains convergent. The improved expression for $w_0$ takes the form
\begin{equation}
        w_0(v) \equiv \dfrac{\sqrt{v-4m^2}}{v^{5/2}}\mathrm{NN}_\sigma\left(\frac{4m^2}{v}\right),\quad v\geq 4m^2\,.
\end{equation} This new ansatz ensures that the integral in Eq.~\ref{eq:Kappa_eqofmot} is convergent. Indeed, given that $\mathrm{NN}_\sigma(x) \xrightarrow[x \to 0]{} const$, we have
\begin{equation}
    w_0(v)=O\left(\dfrac{1}{v^2}\right), \quad \text{as } v\to\infty.
\end{equation}

The fact that $w_0(v)$ vanishes at $v=4m^2$ is not required by numerical stability, but it enhances the training convergence rate. Let us explore why this is the case.
Recall that both $\im f_0$ and $\re f_0$ can be computed from $w_0$ using Eq.~\eqref{eq:F0_eqofmot}. In addition, the threshold behavior for $\im f_0$ is given by Eq.~\eqref{eq:AsymptoticThreshold}. When $\mu^2$ is sent to $+\infty$, we expect the dual expression of $\im f_0$ to align with the primal one. By comparing Eq.~\eqref{eq:F0_eqofmot} with Eq.~\eqref{eq:AsymptoticThreshold}, it becomes clear that $w_0$ must vanish at the threshold.

To train the neural network we numerically computed the integrals required for $\overline{\mathcal{D}}$ by discretizing the region $s\geq 4m^2$ on a grid of 900 points (see Sections~\ref{sec:grid}). An energy cutoff of $10^8m^2$ was sufficient for the bounds to converge. The dual bounds obtained by the neural optimizer for $\mu^2=\infty$ are shown in orange in Figure~\ref{fig:Dual}.

\subsection{Semi-definite programming}
\label{sec:dual_sdpb}
Next, we will use semidefinite programming to derive a bound on $\overline{\mathcal{D}}$. For this purpose, let us define a new objective functional $\overline{\mathcal{D}}^\text{lin}$ of the new dual variables $\{w_0, \chi_0^\text{IR},\chi_0^\text{UV}\}$
\begin{equation}
    \overline{\mathcal{D}}^\text{lin} = \int^{\mu^2}_{4m^2} \frac{dv}{\varphi(v)} \, n_0 \, \chi^\text{IR}_0(v) + \int^\infty_{\mu^2} \frac{dv}{\varphi(v)} \, n_0 \, \chi^\text{UV}_0(v) \ ,\label{eq:dual_lin}
\end{equation}
subject to semi-positiveness conditions
\begin{equation}
    \begin{pmatrix}
        \chi^\text{IR}_0(v) & \quad w_0(v)/n_0 \\
         w_0(v)/n_0 & \quad \chi^\text{IR}_0(v){-}2\overline{\mu}_0(v)
    \end{pmatrix} \succeq 0
    \quad \text{and} \quad
    \begin{pmatrix}
        \chi^\text{UV}_0(v) & 0 \\
        0 & \chi^\text{UV}_0(v){-}2\overline{\mu}_0(v)
    \end{pmatrix} \succeq 0.
\label{eq:SD_conds}
\end{equation}
Two important properties of the new objective are: (i) It is linear in the dual variables, (ii) it is guaranteed that $\overline{\mathcal{D}}^\text{lin} \geq \overline{\mathcal{D}}$ holds, which can be seen by computing the determinants of the constraint matrices in~\eqref{eq:SD_conds} and forcing them to be positive.

All in all, $\overline{\mathcal{D}}^\text{lin}$ provides us rigorous (but not necessarily optimal unless  $\overline{\mathcal{D}}^\text{lin} {=} \overline{\mathcal{D}}$) bounds on the primal objective, and its minimization is amenable to semidefinite linear programming tools. At last, we can state the linearized optimization problem as
\begin{align}
\min_{\{ w_0,\chi^\text{IR},\chi^\text{UV} \}} \,\, \overline{\mathcal{D}}^\text{lin} \quad &\text{subject to} \quad \mathrm{Eqs.~}\eqref{eq:SD_conds}\,.
\label{eq:min_dlin}
\end{align}
We use SDPB to find the solution numerically; more details on the implementation can be found in Appendix~\ref{sdpb_details_dual}. The resulting dual upper bounds on $\max c_0/(32\pi)$ are presented in Figure~\ref{fig:Dual_extra} for various values of $\mu^2$, and the exclusion regions in $(c_0,c_2)$ plane are presented in Figure~\ref{fig:Dual}.

\subsection{UV-IR interplay}
\label{sec:uv-ir-sens}

The model under study exhibits an interesting UV-IR interplay, which we describe here. 
Firstly, recall that while subtractions naively allow for a non-vanishing $\im f_0(s)$ as $s \to \infty$, we have shown that both the real and imaginary part of $f_0(s)$ are forced to vanish through unitarity. A consequence of this is the existence of an extra dispersive sum rule for $c_0$, given in~\eqref{eq:c0-sumrule}.

By switching on and off the $\kappa_0$ parameter in the dual setup, it becomes evident that the unitarization in the high-energy region directly affects the dual bounds on the low-energy observable $c_0$, as depicted in Figure~\ref{fig:Dual_extra}. In particular, we managed to reach the true upper bound of $2.41$ when $\kappa_0 \neq 0$, and we failed in detecting the same bound in the runs with $\kappa_0 = 0$, for no value of $\mu^2$ up to $10^4m^2$. It suggests that pushing $\mu^2$ to exponentially large energies is needed to capture the bounds correctly.

The impact of the UV is particularly significant for purely elastic amplitudes, which exhibit a universal $1/\log(s)^2$ decay at high energies, given by~\eqref{eq:AsymptoticRegge}. 
This implies that truncating the $c_0$ sum rule at a UV scale $\Lambda^2$ introduces an error of order $\frac1{\log(\Lambda^2)}$, necessitating an exponentially large cutoff $\Lambda^2$ to achieve precision.

The UV-IR interplay motivates the choice of the NN architecture with two inputs~$\{x, \log x\}$
that we use in the next section.

\section{Primal optimization}
\label{sec:Primal}

In the \emph{primal} approach, our task is to construct solutions to the S-matrix bootstrap problem of interest explicitly. This means that we populate the complement of the region excluded through the dual. In addition, this approach allows us to analyze the resulting amplitudes in detail and explore the emergence of resonances, the behavior of phase shifts, etc.

In this section, we describe how we implemented the primal approach via two different methods: 1) We use the NN to parameterize the single discontinuity of the amplitude. 2) We use the $\rho$-ansatz for the amplitude itself.

\subsection{Neural optimizer}
\label{sec:Primal_NN}

We start amplitude from the ansatz~\eqref{eq:def_toy_model} that is manifestly crossing-symmetric and analytic. In this section, our goal is to build 
an amplitude which satisfies elastic unitarity through the non-linear equation~\eqref{eq:full-unit-rho} on the imaginary part of the amplitude $\im f_0$.
After we parametrized $\im f_0$ with a neural network, we seek to equation~\eqref{eq:full-unit-rho} which for convenience we write as
\begin{equation}
    |S_0(s)|^2 =1\,,\quad s\geq 4m^2\;.
    \label{eq:elas-unit-Swave}
\end{equation}
Let us describe the details of our implementation.

\paragraph{Ansatz and parameterization with the NN.}

The grid used to train the NN has an exponentially large UV cutoff $s_{\rm{UV}}=10^{100}m^2$. This is required by the UV-IR interplay of our model described in Section~\ref{sec:uv-ir-sens}.

To parameterize $\im f_0$, we use the neural network depicted in Figure~\ref{fig:NN}, which takes two inputs $x=4m^2/s$ and $\log(x)$. The goal is to correctly capture both the threshold behavior near $x=1$ and the Regge behavior close to $x=4m^2/s_{\rm{UV}}\simeq 10^{-100}$. 
In Section~\ref{sec:Dual_NN}, we noted that incorporating analytical results into the neural network ansatz can greatly enhance training. Here, we explain our implementation of this idea in the primal context.

To start with, since two threshold behaviors are allowed, regular and singular, 
see~\eqref{eq:AsymptoticThreshold},
we introduce two different parameterizations to describe them.
\begin{subequations}
The regular ansatz is given by:
\begin{equation}
    \im f_0(s) = 
        \varphi(s) \mathcal{R}(s) \left(1+\mathrm{CELU}\left( \mathrm{NN}_\sigma(\tfrac{4m^2}s) \right)\right),
\label{eq:dressed_NN_reg}
\end{equation}
and the singular ansatz by:
\begin{equation}
    \im f_0(s) = 
        2\dfrac{\mathcal{R}(s)}{\varphi(s)} \left(1+ \mathrm{CELU}\left(\varphi(s)^2\,\mathrm{NN}_\sigma(\tfrac{4m^2}s) \right)\right)\,.
\label{eq:dressed_NN_sing}
\end{equation}
\end{subequations}
In these equations, $\rm{NN}_\sigma$ is the neural network, $\varphi(s)$ is the phase-space factor, CELU is the activation function defined in~\eqref{eq:CELU_function} and $\mathcal{R}(s)$ is defined as
\begin{equation}
        \mathcal{R}(s) \equiv \dfrac{1}{\left(1-\log(4m^2/s)\right)^2} \xrightarrow[s\to\infty]{} 0.\label{eq:ReggeFunc}
    \end{equation}
Let us decipher the precise form of these equations:
\begin{itemize}
    \item Firstly, the function $1+\mathrm{CELU}$ is positive, as can be checked from the definition~\eqref{eq:CELU_function}. Combined with the positivity of $\varphi(s)$ and ${\cal R}(s)$, this implies that the ansatz has positivity built in: $\im f_0(s)\geq0$.\footnote{One might worry that this prevents $\im f_0(s)$ from vanishing at all, but it does not: thanks to the exponential fall-off of CELU, the function $1+\mathrm{CELU}$ can effectively reach zero at finite distance in its argument for all practical and numerical purposes.}
    \item The leading large $s$ behavior of $\im f_0(s)$ is known to be slowly decreasing as $1/\log^2(s)$, see Eq.~\eqref{eq:AsymptoticRegge}. To help the network capture this behavior, we added by hand the factor of $\mathcal{R}(s)$.
    The shift by 1 in the denominator eliminates a potential spurious singularity at $s 
    \to 4m^2$, sparing the network from having to learn to remove it.
    \item The threshold limits differ for the regular and singular ansatz, but they share a common square-root non-analyticity at $s=4m^2$, which is hard to capture for the network.
    We solve this issue by adding a factor of $\varphi(s)$ and $1/\varphi(s)$ for the regular and singular ansatz, respectively.
\end{itemize}

Lastly, in the singular case, a factor $\varphi(s)^2$ is included inside the CELU's input to ensure that the neural network contribution is sub-leading near the threshold. Expanding Eq.~\eqref{eq:dressed_NN_sing} around $s\to 4m^2$ we obtain
\begin{equation}
    \im f_0(s) = \dfrac{2}{\varphi(s)}-4\varphi(s)+2\varphi(s)\mathrm{NN}_\sigma(1)+O(\varphi(s)^2), 
\end{equation}
which reproduces the threshold expansion expected from elastic unitarity, see e.g. Eq.~(5.6) in~\cite{Correia:2020xtr}.

\paragraph{Loss function.}
The training of the NN seeks to minimize the loss; therefore, we want to choose a loss that is minimized when elastic unitarity is achieved. Different choices for the loss are possible. We choose the simple option $(\rm{LHS}-\rm{RHS})^2$ of Eq.~\ref{eq:elas-unit-Swave}: this choice does not distinguish between inelasticities ($|S_0|<1$) and unitarity violation ($|S_0|>1$) but for the problem at hand, this fact will prove harmless.
The exact expression of the loss we used is given by
\begin{equation}
    \mathcal{L}_0=\dfrac{1}{N}\!\sum_{\substack{s\in\mathcal{S}_N}}\!\!\dfrac{\Big(|S_0(s)|^2-1\Big)^2}{\mathcal{R}^{1/2}(s)}.\label{eq:loss_U}
\end{equation}
The factor $\mathcal{R}^{-1/2}(s)$ assigns more weight to the high-energy points and empirically helps the training to capture the correct UV behavior.%
\footnote{Intuititively, this can be understood as follows. At high energies, in the logarithmically decaying regime, $f_0\to 0$ and leads to $ (|S_0(s)|^2-1)^2\underset{s\to\infty}{\sim} \dfrac{\mathrm{const}}{\log^4(s)}$. The factor of ${\cal R}^{-1/2}\sim \log(s) $ helps to redistribute the weight in this tail.}

The grid $\mathcal{S}_N$ used for $s$ consists of $800$ points, with $300$ of them log-spaced between $m^2$ and $10^{100}m^2$, see also Appendix~\ref{sec:grid}. The integration over energies in the dispersion relations hidden in $S_0$ (see~\eqref{eq:ltot-def}) is reduced to matrix multiplication~(see Appendix~\ref{sec:integration_matrix} for the details.

With the definitions~(\ref{eq:dressed_NN_reg},~\ref{eq:dressed_NN_sing}) for $\im f_0(s)$ and \eqref{eq:loss_U} for the loss function we can enforce unitarity for various values of $c_0$ which is a free parameter in this set-up. Our goal at this point is to construct amplitudes anywhere inside the region derived by the dual method. Therefore, we need not only to fix $c_0$ but also $c_2$.
To achieve this, we simply added an extra term $\mathcal{L}_{c_2}$ to the loss function, defined by
\begin{equation}
    \mathcal{L}_{c_2} = w_2\left(c_2^\mathrm{target}-16\!\!\int_{4m^2}^{\infty}\!\!\!\!\mathrm{d}v\,\dfrac{\im f_0(v)}{(v-4m^2/3)^3}\right)^2,\label{eq:loss_c2}
\end{equation}
so that our total loss becomes
\begin{equation}
\label{eq:ltot-def}
\mathcal{L}_{\mathrm{tot}}=\mathcal{L}_0+\mathcal{L}_{c_2} 
 .
\end{equation} 
In \eqref{eq:loss_c2}, $c_2^\mathrm{target}$ is a target value for $c_2$ and $w_2$ a possible weight factor. One can check that $\mathcal{L}_{c_2}$ is minimized when $c_2=c_2^\mathrm{target}$ using the definition~\eqref{eq:c2-sumrule}. The weight factor $w_2$ is critical to ensuring that the two losses, $\mathcal{L}_0$ and $\mathcal{L}_{c_2}$, are of comparable magnitude. If one term dominates the other, the neural optimizer might focus on minimizing only the dominant term. In such a case, only the corresponding constraint (associated with either $\mathcal{L}_0$ or $\mathcal{L}_{c_2}$) would be enforced during training, potentially neglecting the other. These weight factors must be chosen appropriately in a given problem. We ultimately used the value $w_2=1$.

\paragraph{Primal boundary.}
Using the loss $\mathcal{L}_\mathrm{tot}$ we can thus train the neural network to solve~\eqref{eq:elas-unit-Swave} for any pair $(c_0,c_2)$. The dual approach provides the region in the $(c_0,c_2)$ plane where \textbf{ACU} amplitudes are not excluded. 

Because of the procedure that we adopted, based on minimizing a loss, nothing prevents the network's training from giving results in the unphysical, excluded region. Therefore, the question is: How do we construct the physical region in the primal approach? 
It turns out that amplitudes deep in the excluded regions have a clear pathological behavior: their loss remains of order 1 or worse, and unitarity is clearly not satisfied.

Likewise, inside the allowed region, the loss converges to a small value at the end of the training, consistently achieving $\mathcal{L}_\mathrm{tot}<10^{-5}$.
The square root of the loss $\sqrt{\mathcal{L}_\mathrm{tot}}$, can be thought of as an approximate measure of the deviation of $|S_0|$ from unity at each grid point
\begin{equation}
    |S_0|\simeq 1 \pm \sqrt{\mathcal{L}_\mathrm{tot}}\;.
\end{equation}

The more subtle question is how to identify the precise boundary of the allowed region.
\begin{figure}
    \centering
    \hspace{-2.2cm}
    \includegraphics[]{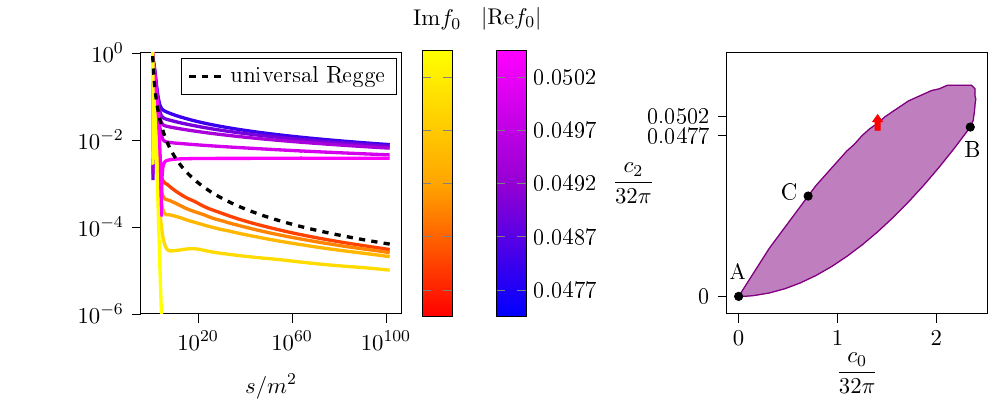}
    \caption{\underline{Left.} The Regge behavior of $f_0(s)$ for various values of $c_2$ as the boundary of the allowed space is crossed. $\frac{c_0}{32\pi}$ is fixed at 1.4. The imaginary part $\im f_0(s)$ is shown in Red-to-Yellow as $c_2$ increases while the real part $|\re f_0(s)|$ is depicted in Blue-to-Magenta as $c_2$ increases. The last $|\re f_0(s)|$ curve (for $c_2=0.0502$, shown in magenta) increases with energy violating~\eqref{eq:ReggeLimits}.
     The Regge behavior $\frac{2\pi^2}{9\log^2 s}$ is represented by the dashed black line. \underline{Right.} The corresponding positions of the amplitudes within the allowed space.}
    \label{fig:reggeL45}
\end{figure}
One might attempt to do so by inspecting unitarity violations after training. Intuitively, these violations should increase as we move out of the allowed region in the $(c_0,c_2)$-plane. 
This approach correctly identifies the boundaries A-B and A-C, shown in the right plot of Figure~\ref{fig:reggeL45}.
However, it fails along the B-C boundary. If we used this criterion, we would obtain a wrong upper bound, $\frac{c_0}{32\pi}=2.66$ (which is the upper bound of the full problem) instead of $2.41$.

We discovered that the criterion $f_0(s) \xrightarrow[s\to\infty]{}0$ is what accurately determines the boundary. 
Specifically, just as we cross inside the region excluded region, $f_0(s)$ no longer vanishes as it should in the Regge limit; instead, $\re f_0$ or $\im f_0$ appears to diverge logarithmically. An example of this behavior is given in Figure~\ref{fig:reggeL45}. The left plot illustrates the Regge behavior of several amplitudes at fixed $c_0$ for increasing $c_2$. The real part $|\re f_0(s)|$ is shown in a blue-to-magenta gradient and $\im f_0(s)$ in a red-to-yellow gradient. The right plot indicates the location of the amplitudes depicted in the left plot in the $(c_0,\, c_2)$ plane. 
The definition of our criterion is therefore that the boundary occurs at the maximal value of $c_2$ (or $c_0$, if moving horizontally) where both $\im f_0(s)$ and $\re f_0$ continue to decrease in the Regge limit, i.e., $\dfrac{c_2}{32\pi}=0.0497$ in this example.

In conclusion, using the expression~(\ref{eq:dressed_NN_reg}~\ref{eq:dressed_NN_sing}) and the total loss $\mathcal{L}_\mathrm{tot}=\mathcal{L}_0 + \mathcal{L}_{c_2}$, we can train the neural network for any pair $(c_0,\, c_2)$. Inside the allowed region can we achieve losses $\mathcal{L}_\mathrm{tot}<10^{-5}$.

Moving outside of the allowed region, the convergence of the loss does not necessarily degrade -- this is the case for the boundary between B and C. However, beyond the allowed region, the Regge limit of the amplitude is no longer consistent with unitarity, which our choice of loss $\mathcal{L}_\mathrm{tot}$ fails to capture. The expected Regge behavior of our amplitude, characterized by its asymptotic decay, emerges only at very high energies. This is why we need an exponentially high-energy cutoff $s_\mathrm{UV}=10^{100}m^2$ to verify it.

\subsection{The $\rho$-bootstrap}
\label{sec:rho_bootstrap}

We can alternatively solve the primal extremization problem in the $(c_0,c_2)$ plane by parameterizing $T(s,t)$ via a suitable basis of functions, the $\rho$-basis, as in~\cite{Paulos:2017fhb}. More precisely, we use the wavelet basis introduced in \cite{EliasMiro:2022xaa}:
\begin{equation}
    \label{eq:rho-map}
    \rho_\sigma(s) = \frac{\sqrt{\sigma-4m^2}-\sqrt{4m^2-s}}{\sqrt{\sigma-4m^2}+\sqrt{4m^2-s}} \ ,
\end{equation}
where $\rho_\sigma(s)$ maps the principal sheet in the complex $s$-plane to a unit disk, and $\sigma$ controls the position of its center.\footnote{Points $s = \{4,\sigma,8{-}\sigma,\infty\}$ respectively map to $\rho=\{1,i,0,-1\}$}

By assumption we set the double discontinuity to zero, which leads to the following ansatz
\begin{equation}
    T(s,t,u) = \sum_{\sigma \in \Sigma_N} \alpha_\sigma \Big( \rho_\sigma(s) + \rho_\sigma(t) + \rho_\sigma(u) \Big) \ .
    \label{eq:primal_ansatz}
\end{equation}
The ansatz \eqref{eq:primal_ansatz} is crossing-symmetric by construction, and it develops a single discontinuity whenever $s,t,u>4m^2$. The wavelet parameters $\sigma$ are chosen from the set $\Sigma_N$ described in the Appendix~\ref{sdpb_details_primal}. 
This ansatz is derived from the full problem by omitting double product terms like $\rho_\sigma(s) \rho_\tau(t)$, which naturally eliminates any potential double discontinuity.\footnote{Curiously, a similar, double discontinuity-free ansatz was also used long ago in~\cite[Section~2]{Auberson:1977ss}. They obtained primal optimal values for $c_0$ close to our optimal bounds, namely, they found amplitudes with $\frac{c_0}{32\pi}=2.38$.}

We then look for solutions in the space spanned by the real coefficients $\{\alpha_\sigma\}$ which obey the unitarity condition \eqref{eq:S0_uni} imposed via $\mathcal{U}_0 \succeq 0$ as in \eqref{eq:unitarity_matrix}. The benefit of using $\mathcal{U}_0$ is that it is linear in the unknowns $\{\alpha_\sigma\}$, thus it allows us to formulate a semi-definite linear optimization problem. In the case of the $c_0$ maximization, we can state it as 
\begin{align}
\max_{\{ \alpha_\sigma \}} 
\, c_0 \quad &\text{subject to} \quad \mathcal{U}_0(s) \succeq 0, ~~~s \geq 4m^2 \ . \label{eq:primal_sdlp}
\end{align}
We use SDPB to find the solution numerically; more details on the implementation can be found in Appendix~\ref{sdpb_details_primal}.
The primal search with the $\rho$-basis in $(c_0,c_2)$ plane agrees with the NN results completely, and the results are shown in Figure~\ref{fig:results}.


\section{Discussion}
\label{sec:Discussion}

In the previous sections, we have described the neural optimizer and the traditional nonperturbative bootstrap to characterize the space of amplitudes with zero double discontinuity. 
Although the problem we applied our machinery to is an intermediate step to solving the full bootstrap problem, where unitarity is enforced in all partial waves, the amplitudes that we have constructed have many interesting properties. In this section, we discuss these properties, as well as more general lessons learned along the way. 

\subsection{Physics of low-spin dominated amplitudes}
\label{sec:low-spin}
The amplitudes constructed in the paper provide an example of what we can call low-spin dominated amplitudes. By this, we mean that scattering mostly takes place in the S-wave, with all the higher spin partial waves being relatively small.\footnote{Our amplitudes are not complete in the sense that they violate unitarity for partial waves with $J>0$; however, as we discuss further below, we expect that this can be easily fixed.} 
Examples of such amplitudes include weakly coupled $\phi^4$ theory or the $O(N)$ model at large $N$ \cite{Carmi:2018qzm,Henning:2022xlj}.

Let us describe the S-matrices along the boundary of the single-disc almond. They are characterized not only by their threshold behavior and high-energy decay but also by the positions of resonances. A spin-$J$ resonance corresponds to a zero of the partial wave, $S_J(s_*)=0$, in the upper-half plane of the principal sheet. As usual, through real analyticity 
\begin{equation}
\label{eq:real-analyticity}
S_J({s}^*) = {S_J(s)}^* \, ,
\end{equation}
there is always an associated complex conjugate zero in the lower-half plane of the principal sheet.
In this work, all resonances occur in the spin $J=0$ partial wave. These features are dynamically generated by the optimizers as an output of the procedure.

\begin{figure}
    \hspace{-1.2cm}
    \includegraphics[]{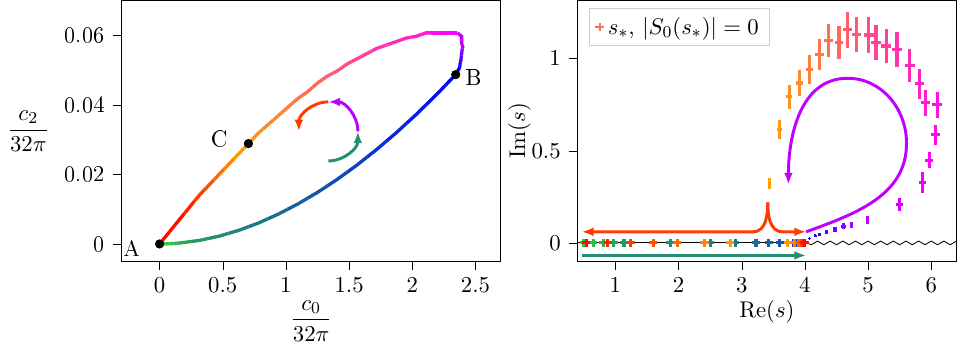}
    \caption{\underline{Left.} 
    Decomposition of the single-disc almond boundary into three segments, according to the resonance pattern.
    Segment A-B agrees with the boundary of the full allowed space shown in green in Figures~\ref{fig:full_region}.
    \underline{Right.} Position of the resonance in the complex $s$ plane as we move along the boundary of the allowed region. We only show the upper half-plane, which is mapped to the lower half-plane through real analyticity~Eq.\eqref{eq:real-analyticity}. The error bars come from the grid points used to detect the vanishing of $|S_0|$ in the complex $x=4m^2/s$ plane. On the real axis, we remove the error bars to improve readability.}
    \label{fig:Zeros}
\end{figure}

Below, we describe the precise pattern in which these zeros appear and move, as we go around the boundary. We have identified three special points A-B-C and three continuously connected regions between them, as depicted in Figure~\ref{fig:Zeros}.

\paragraph{Lower boundary A-B.}

Segment A-B is the lower boundary of the single-disc almond. This boundary coincides with the boundary of the full problem solved in~\cite{EliasMiro:2022xaa,Chen:2022nym} depicted in green in Figure~\ref{fig:mainres}, it would be interesting to understand why in detail. 

Along the lower boundary, we found empirically that only the regular threshold boundary condition is admissible. 

\begin{figure}
    \centering
    \begin{subfigure}[t]{0.32\textwidth}
    \includegraphics[scale=0.92]{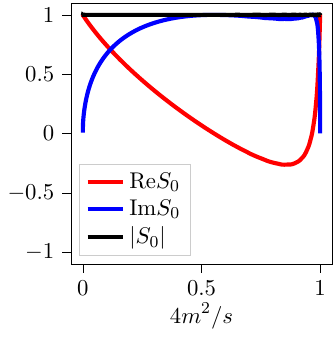}
    \caption{Neural optimizer:\\A-B boundary\\ $\frac{c_0}{32\pi} = 1.9$\\}
    \label{fig:nnA-B}
    \end{subfigure}
\hfill
\begin{subfigure}[t]{0.32\textwidth}
    \includegraphics[scale=0.92]{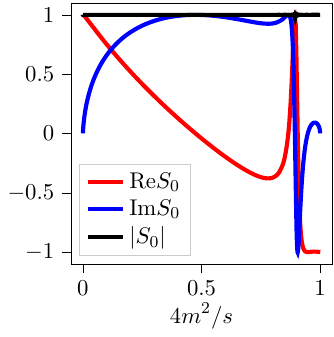}
    \caption{Neural optimizer:\\B-C boundary\\ at maximal value $\frac{c_0}{32\pi} = 2.4$.\\}
    \label{fig:nnB-C}
\end{subfigure}
\hfill
\begin{subfigure}[t]{0.32\textwidth}
    \includegraphics[scale=0.92]{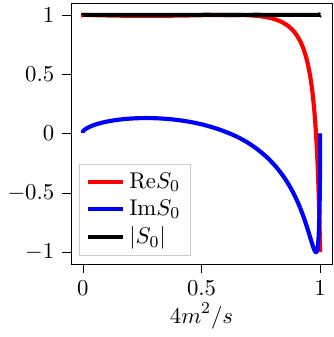}
    \caption{Neural optimizer:\\A-C boundary\\ $\frac{c_0}{32\pi} = 0.47$\\}
    \label{fig:nnA-C}
\end{subfigure}

\vspace{1em} 

\begin{subfigure}[t]{0.32\textwidth}
    \includegraphics[scale=0.92]{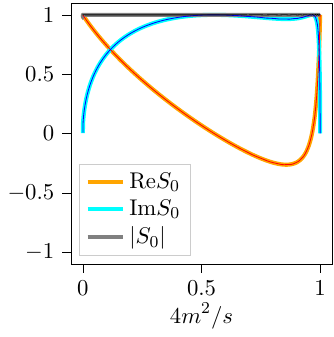}
    \caption{$\rho$-bootstrap:\\A-B boundary\\ $\frac{c_0}{32\pi} = 1.9$}
    \label{fig:bA-B}
    \end{subfigure}
\hfill
\begin{subfigure}[t]{0.32\textwidth}
    \includegraphics[scale=0.92]{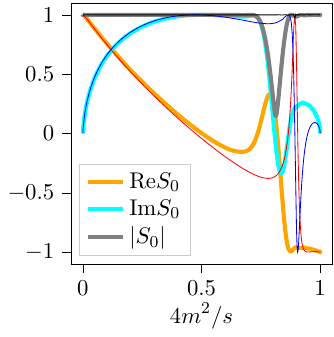}
    \caption{$\rho$-bootstrap:\\B-C boundary\\ at maximal value $\frac{c_0}{32\pi} = 2.4$.}
    \label{fig:bB-C}
\end{subfigure}
\hfill
\begin{subfigure}[t]{0.32\textwidth}
    \includegraphics[scale=0.92]{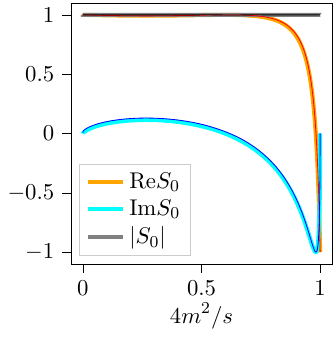}
    \caption{$\rho$-bootstrap:\\A-C boundary\\ $\frac{c_0}{32\pi} = 0.47$}
    \label{fig:bA-C}
\end{subfigure}

\caption{Examples of the $S_0$ partial wave obtained on the boundary of the single-disc almond, $\re S_0$, $\im S_0$ and $|S_0|$ are plotted as a function of $4m^2/s$. The \underline{top} row shows partial wave obtained by the NN at different locations on the boundary. The \underline{bottom} row shows the same amplitude obtained by the $\rho$-bootstrap method; the NN result from the row above is recalled in thin lines. Both methods produce the same partial wave unless it exhibits a resonance (B-C boundary), in this case, the $\rho$-bootstrap generates inelasticity, as can be seen from the large grey downward spike in Figure~\ref{fig:nnB-C}.}
\end{figure}

The S-matrices $S_0(s)$ on the A-B boundary have exactly one zero, which is a real zero below the threshold, see Figure~\ref{fig:Zeros}. 
As $c_0$ increases, the zero moves towards the threshold. A sub-threshold zero introduces a shift by $\pi$ in the phase of the amplitude, $\frac1{i}(\log S_0(s))$, in comparison to an above-threshold zero, for which the shift is by $2\pi$. Examples of both cases can be seen in Figures~\ref{fig:nnA-B}~and~\ref{fig:nnB-C}, respectively.

However, the zero does not reach the threshold at point B: instead, it would reach the threshold for the full problem, if we continued along the lower boundary to the maximal coupling $2.66$. At point $B$, the zero is very close but not exactly {at} the threshold.\footnote{The data from~\cite{Chen:2022nym,EliasMiro:2022xaa} clearly shows that $4m^2$ is reached only at $c_0=2.66$. This is consistent with the extrapolation of our data, which follows a power law.}

The emergence of zero near point A can be demonstrated perturbatively. Consider the asymptotically free $\lambda \phi^4$ amplitude (it is defined by $\lambda < 0$~\cite{Symanzik:1961}). Below the threshold, at leading order in $\lambda$, we have
\begin{equation}
    \re f_0(s) = -\lambda/(16\pi) + O(\lambda^2) \quad , \quad \im f_0(s) = 0 \quad , \quad s \in [0,4m^2] \, .
\end{equation}
The zero of the partial amplitude is then given by the solution to
\begin{equation}
    S_0(s_*) = 0 \quad \Leftrightarrow \quad 1 + \sqrt{\frac{4m^2-s_*}{s_*}} \frac{\lambda}{16\pi} = 0 \quad , \quad \lambda < 0 \, ,
    \label{eq:S0_pertzero}
\end{equation}
which gives $s_*=\lambda^2m^2/(8\pi)^2+O(\lambda^2)$. This confirms the presence of a zero parametrically close to the origin in the $s$-plane, as we approach point A from the lower boundary, since $c_0=-\lambda+O(\lambda^2)$.\footnote{Remark that for the opposite sign theory ($\lambda>0$) there is no solution for $s_*$ to~\eqref{eq:S0_pertzero} below threshold. We cannot access this regime with our toy model because of the sum rule \eqref{eq:c0-sumrule} that forces $c_0>0$.}

In general, we were not able to generate consistent solutions with regular threshold behavior inside the single-disc almond, except for the A-B segment. 
Inside the single-disc almond, a typical solution has singular threshold behavior and a resonance somewhere in the complex plane.

As we move vertically in the $(c_0,c_2)$ plane, towards the lower boundary, at fixed $c_0$, the resonance and its complex conjugate travel towards some location on the real line, above the threshold, $s>4m^2$. The transition to amplitudes \textit{on} the boundary appears discontinuous: the resonance and the complex conjugate disappear, the threshold behavior
becomes regular, and the aforementioned zero appears on the real line, \textit{below} the threshold.

\begin{figure}
    \centering
    \includegraphics[]{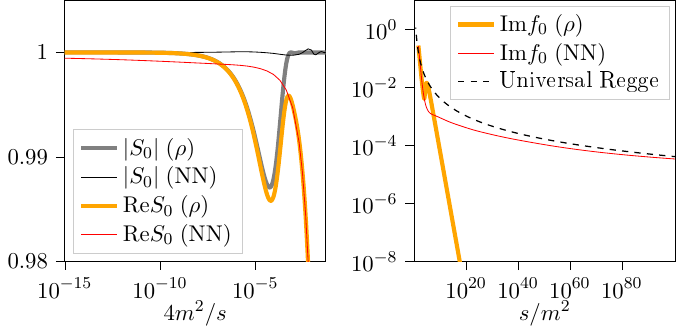}
    \caption{\underline{Right.} Zoom on high energy of Figure~\ref{fig:bA-B}. We observe in this example that $\rho$-bootstrap generates inelasticity (see $|S_0|<1$ in gray). The small oscillation of $|S_0|$ obtained by the NN are visible in black in this plot, we will comment on these oscillation in Section~\ref{sec:validating_NN}.  \underline{Left.} Regge behavior of $\mathrm{Im} f_0$ obtained both from $\rho$-bootstrap and NN. The partial wave obtained by the NN (in red) follows the universal Regge behavior Eq.~\eqref{eq:AsymptoticRegge} (in dashed black). The $\rho$-bootstrap partial wave decreases as a power law.}
    \label{fig:Regge_NN_vs_Rho}
\end{figure}

\paragraph{Upper boundary B-C}
The point B marks the discontinuous transition between regular and singular threshold behavior. The same type of transition as described above occurs, and a pair of complex conjugate zeros appears at a location close to $s=4m^2$.

On the B-C boundary, the S-matrices are, therefore, identical to the generic amplitudes in the bulk of the single-disc almond and have two complex conjugate zeros.

As we move along B-C towards C, the zero and its complex conjugate wander in the complex plane, following a teardrop shape (see Figure~\ref{fig:Zeros}). 
\paragraph{Upper boundary C-A}
At point C, the two conjugate zeros meet on the real axis and give rise to two real zeros. As we continue the journey back to the origin point A, the pair of zeros move towards the origin and the threshold, respectively.
At point A, these zeros touch these two extremities, and the threshold singularity disappears, thereby smoothly connecting back to the {weakly coupled} theory.
A typical S-matrix on this branch is displayed in Figure~\ref{fig:nnA-B}.

In conclusion, this pattern of resonances appears nontrivial and seems related to what is observed in the full problem~\cite{Chen:2022nym,Gumus:2023xbs}. It would be interesting to re-investigate the question and test this picture when we study the full problem. For instance, it would be interesting to know if, for the full problem, regular boundary conditions are constrained to some small region, as for the amplitudes with no double discontinuity, or if they can be generically achieved in the bulk of the allowed space.

\subsection{Unitarity violation in $J>0$ partial waves}
\label{sec:unit-viol}

Even though our model only satisfies S-wave unitarity, the higher spin partial waves~$J>0$, are necessarily present in the full amplitude as a consequence of crossing.

\begin{figure}
    \centering
    \includegraphics[]{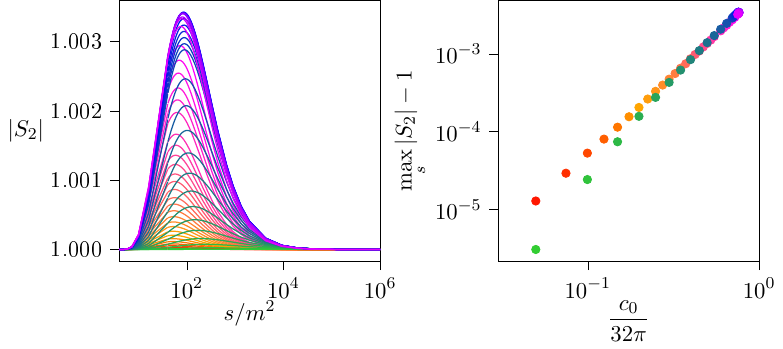}
    \caption{Violation of unitarity in the spin-2 partial wave. The colors match the ones of Figure~\ref{fig:Zeros} and indicate the position around the single-disc almond. \underline{Left.} $|S_2|$ as a function of energy for several amplitudes around the single-disc almond. \underline{Right.} maximal unitarity violation ($\max_s|S_2|-1$) as function of $c_0$.}
    \label{fig:S2}
\end{figure}

The absence of double discontinuity, however, renders all the higher partial waves non-unitarity; see discussion around Eq.~\eqref{eq:non-unit}. In Figure~\ref{fig:S2}, we show explicitly the size of unitarity violation in the $J=2$ partial wave for the amplitudes along the boundary of the single-disc almond. The violation is small, and we expect that adding a small double discontinuity will make them unitarity and, thus, potentially physical. 

In the previous work~\cite{Tourkine:2023xtu}, we achieved something related. In the small region where the fixed point algorithm converged (see the red line in Figure~\ref{fig:results}), we saw precisely that the double-discontinuity precisely unitarizes the higher waves without modifying much the $J=0$ wave.

In order to know if the same phenomenon happens inside our whole single-disc almond, we need to develop our neural optimization solver for the full problem. If the answer happens to be yes, the single-disc almond would then correspond to a region where $S_0$ and the higher partial waves can essentially be decoupled and unitarized separately. This remains conjectural.

What we can say without conjecture is that in the \textit{complement} of the single-disc almond, within the space of fully unitary, nonperturbative S-matrices, $S_0$ \textit{cannot} be unitarized with a zero double discontinuity. This {gives an interesting new piece of information} on the role and need of double discontinuity in the solution to \textbf{ACU}.

\section{Numerical aspects}
\label{sec:numerical}
In this section, we summarize and discuss the numerical aspects of this work. First, we recapitulate how our results validate the NN methodology, emphasizing that it allows us to make the Atkinson-Mandelstam approach to the S-matrix bootstrap a practical tool for exploring the space of scattering amplitudes. Second, we compare the NN approach to other non-linear bootstrap schemes, such as fixed-point iterations and Newton's method.

\subsection{Validating the neural optimizer}
\label{sec:validating_NN}
\paragraph{Primal strategy.}
With the methodology described in the text, both the $\rho$-bootstrap and NN approaches yield the same allowed region for our model, which is summarized in Figure~\ref{fig:mainres} in the introduction. This is the main point that validates the NN method. Let us delve a bit more into this statement and reflect upon the differences between the two methods.

Comparing the pros and cons of each method a priori is challenging because they tackle the bootstrap problem from two different angles.
Both methods begin with an ansatz satisfying analyticity and crossing symmetry.
The NN approach approximately enforces unitarity by solving a non-linear equation for the discontinuity of the amplitude, $$|S_0|^2=1.$$In contrast, the $\rho$-bootstrap imposes unitarity as a strict constraint $$|S_0|\leq 1 \ , $$ while solving a linear optimization problem. After optimization, the $\rho$-bootstrap typically produces solutions where the unitarity constraint is saturated at all energies, $|S_0|=1$.
This fact makes it possible to compare the performance of both methods.

The first notable difference is that unitarity ($|S_0| \leq 1$) is exactly satisfied by the $\rho$-bootstrap on the grid points, while the NN allows for small violations, typically within $10^{-3}$ of unity, even on grid points. This is because the NN simply tries to minimize the loss \eqref{eq:loss_U} and does not distinguish positive violations of unitarity and small inelasticity. These violations decrease as the number of layers and the number of grid points and appear harmless for our problem.

The emerging resonances are better described by the NN. The $\rho$-bootstrap often generates significant inelasticity around resonances, even within the elastic region ($4m^2\leq s\leq 16m^2$), whereas the NN maintains $|S_0|$ closer to 1, providing a more accurate representation, see Figure~\ref{fig:bB-C}. This presumably comes from the greater flexibility of the NN being a universal approximator and the difficulty of the $\rho$-parametrization to accommodate fast oscillating solutions at finite $N_{max}$. We expect this inelasticity to go to zero as $N_{max}\to\infty$, but this comes at a great computational cost.

As for the high energy limit, the NN captures the true Regge behavior imposed by elastic unitarity, as shown in Figure~\ref{fig:reggeL45}. The $\rho$-bootstrap struggles in this regime to stay elastic, and it introduces a power-like decay after $s \simeq 10^{5} m^2$ accompanied by a significant inelasticity, see Figure~\ref{fig:Regge_NN_vs_Rho}. 
This is another instance where the NN proves flexibility to parametrize exponentially large energies faithfully, which was crucial to obtain correct bounds as explained in Section~\ref{sec:uv-ir-sens}.

Staying at the level of the primal approach, the definition of the allowed space is more straightforward in the $\rho$-bootstrap: the linear optimization algorithm maximizes or minimizes a target quantity (e.g., $c_0$, $c_2$, etc.) under the unitarity constraint, immediately providing the boundary beyond which unitarity breaks down. For the NN, defining the boundary requires a more nuanced approach. The criterion used to delineate the boundary was explained in Section~\ref{sec:Primal_NN}.

\paragraph{Dual strategy.}
From the dual approach perspective, the neural optimizer and semi-definite dual bootstrap yield identical bounds. The only technical difference is that the NN does not need the extra linearization step. It would be interesting to see if there are nontrivial consequences of this fact in other contexts.

\paragraph{Computational costs.}
In both primal and dual setups, SDPB spends $O(10)$ minutes on a machine with 32 cores to produce an extremal amplitude, that is a single point on the boundary of the single-disc almond Figure~\ref{fig:results}. 
In contrast, the dual NN requires approximately $O(10)$ minutes to converge and provide a single point on the boundary when running on 4 cores on a standard laptop. For the primal NN, $O(10)$ minutes are needed to construct a single amplitude for a fixed $(c_0,c_2)$ coordinate. Detecting the boundary requires constructing amplitudes at multiple points, typically ten. Thus, finding a point on the boundary takes a total of $O(100)$ minutes using the same setup.

\subsection{Making Atkinson-Mandelstam practical}
\label{eq:making-Atkinson}

A second crucial point, mentioned as early as the abstract, is that the machine-learning approach makes the Atkinson-Mandelstam method not just a mechanism to construct precise amplitudes but an actual, practical tool to explore the space of amplitudes. 

By default, the iterative strategy initially put forward by Mandelstam~\cite{Mandelstam:1958xc,Mandelstam:1959bc,Mandelstam:1963iyb} suffers from an important drawback: it is formulated with an a priori unknown inelasticity input and no effective mechanism to explore the space of inelasticities is given.
Even though we focused on fully elastic scattering in this paper, extending the neural optimizer setup to include inelasticity poses no fundamental difficulty. This can be achieved by replacing the loss $\mathcal{L}_0$ \eqref{eq:loss_U} with $\mathcal{L}_{\rm{el}}+\mathcal{L}_{\rm{inel}}$, where
\begin{equation}
    \mathcal{L}_\mathrm{el}=\dfrac{1}{N}\!\sum_{\substack{s\in\mathcal{S}_N\\s \leq s_{\rm{MP}}}}\!\!\dfrac{\Big(|S_0(s)|^2-1\Big)^2}{\mathcal{R}(s)},
\end{equation}
ensures elastic unitarity below the multi-particle threshold $s_{\rm{MP}}$ in the same way as before, and
\begin{equation}
    \label{eq:uniineq}
    \mathcal{L}_\mathrm{inel}=w_\mathrm{inel}\left(1-\max_{\substack{s\in\mathcal{S}_N\\s \geq s_{\rm{MP}}}}|S_0(s)|^2\right)^2,
\end{equation}
ensures that $|S_0|\leq 1$ in the inelastic region.
%
The weight $w_\mathrm{inel}$ plays a similar role as $w_2$ in Eq.~\eqref{eq:loss_c2}. 

A noteworthy feature of this approach is that the inelasticity is generated by the neural network itself. It does not require explicit modeling of the inelastic region as is the case for the standard iterative Atkinson-Mandelstam paradigm. It would be interesting to study the inelastic case in future work, once we incorporate the double discontinuity into our neural optimizer.

Another drawback of the iterative procedure is that it makes it cumbersome to fix Taylor coefficients and, thus, hard to navigate in the allowed space thereof. It can, in principle, be done, but it requires adding more subtractions. For instance, in our set-up, one subtraction allowed to fix $c_0$. In order to fix $c_2$, we would need two more (see the discussion in footnote~\ref{fn:unsubtractions}), and implementing it with the iterative strategy would require having to recompute all the numerical integrals. 
We saw that the neural optimizer allows for a much more flexible approach: it is sufficient to add the new loss term $\mathcal{L}_{c_2}$ to guide the network to a target value of $c_2$. 

This can be readily generalized to incorporate additional target features, such as resonances, Adler zeros, and inelasticities. Such generalizations will prove essential for further developing the Atkinson-Mandelstam program.

\subsection{Comparison with fixed-point iterations and Newton's method}

Let us now compare the Neural Optimizer to Mandelstam and Atkinson's original, non-linear, iterative strategies: the fixed-point method and Newton's method. 
We will see that these two methods face convergence issues that the neural network avoids.

Before that, let us just recall that, and as reviewed in~\cite{Tourkine:2023xtu}, the iterative approach, originally suggested by Mandelstam, was used by Atkinson to rigorously prove, by means of nonlinear functional analysis, the existence and uniqueness of amplitudes satisfying crossing, maximal analyticity, elastic and inelastic unitarity.
To this day, these theorems by Atkinson are some of the few rigorous, constructive results in the S-matrix bootstrap program.

\paragraph{Fixed point iteration.}
The equation~\eqref{eq:full-unit-rho} can be written as
\begin{equation}
    \im f_0(x) = \Phi(\im f_0,x),\label{eq:functional}
\end{equation}
where $\Phi$ is a non-linear functional with a unique free parameter: $c_0$. A standard approach to solve such non-linear functional equations is via fixed-point iteration: starting from an initial guess $g_{(0)}$ for $\im f_0$, one defines the iterative sequence
\begin{equation}
g_{(n+1)}(x) = \Phi(g_{(n)}, x).
\end{equation}
If the fixed-point is attractive, and if one starts the iteration ``close enough" from it, the sequence $(g_{(n)})_{n\in\mathbb{N}}$ converges to the solution $g_{(\infty)}\equiv\im f_0$ that solves \eqref{eq:functional}.

This precise equation was already studied using iterative method in the previous work~\cite{Tourkine:2023xtu}. However, as witnessed there, its applicability is limited: the iteration only converges for a restricted range of the parameter $c_0$, which does not cover at all the entire primal space. 

The region accessible with this method is shown in red in Figure~\ref{fig:results}. It happens to lie on the lower boundary because only the regular boundary condition yields a non-repulsive fixed-point, as was observed in~\cite{Tourkine:2023xtu} and in this work.

\paragraph{Newton's method.}

In order to extend the range of convergence of the fixed-point iteration, a standard idea, suggested by Atkinson~\cite{Atkinson:1970zza} and used in $d=2$ in~\cite{Tourkine:2021fqh}, is to use Newton's method, and look for the roots of Eq.~\eqref{eq:functional}, rewritten as
\begin{equation}
    \im f_0(x) - \Phi(\im f_0,x)=0\,.
    \label{eq:newton}
\end{equation}
Newton's method requires using the Jacobian
\begin{equation}
    J_\Phi(x,y) = \dfrac{\partial\Phi(\im f_0,x)}{\partial\im f_0(y)}.
\end{equation}
and is implemented by solving the following integral equation at each iteration:
\begin{equation}
    \int(\mathbb{1}-J_\Phi(x,y))(g_{(n+1)}(y)-g_{(n)}(y))\mathrm{d}y = \Phi(g_{(n)},x) - g_{(n)}(x).\label{eq:Newton}
\end{equation}
Discretizing $x$ and $y$ on a finite grid of points $0=x_0\leq x_1\leq\cdots\leq x_N =1$, allows the integral in Eq.~\ref{eq:Newton} to be transformed into a matrix multiplication.
The iterative sequence $(g_{(n)})_{n\in\mathbb{N}}$ will converge to the solution provided two conditions are met: (i) the initial guess must be sufficiently close to the true solution $\im f_0$, and (ii)  the Jacobian $J_\Phi(x,y)$ must have no eigenvalue equal to 1, which otherwise would lead to a singularity in Eq.~\eqref{eq:Newton}.

For the purpose of this comparison, we implemented Newton's method as described. Starting from a constant $g_{(0)}$, the algorithm converges to a solution that lies on the lower boundary of the allowed space shown in Figure~\ref{fig:results}. 
The convergence region can be expanded by initializing the algorithm with solutions at smaller $c_0$ and slowly increasing this value. This process works until the Jacobian becomes singular. 
The final convergence region is shown in orange in Figure~\ref{fig:results}. For $\frac{c_0}{32\pi}\gtrsim 1.6$ the algorithm diverges, even if hot-started \textit{on} the actual solution.

To explore whether convergence could be achieved inside the single-disc almond, we attempted to hot-start Newton's method by initializing $f_0$ with the neural network's output. We tested this approach across various amplitudes and $(c_0,c_2)$ values. In some cases, the method converged, but in others, it did not. Despite extensive experimentation, we could not identify a relationship between the initial conditions and convergence success.

Therefore, the neural optimizer was superior to Newton's method in terms of convergence range. However, Newton's method is advantageous in terms of precision, and it would be interesting to imagine a scheme where we use both the neural optimizer and then Newton's method, hot-started on it, to improve the precision of our solutions.
In~\cite{Ebel:2024svb,Ebel:2024nof}, a technique was developped to deal with problematic eigenvalues in the Newton's method Jacobian. It would be interesting to see if a similar technique could be developped for our problem.

\section{Future directions}
\label{sec:FutureDirections}

The obvious next step is to apply neural optimizers to the physical scattering amplitudes without a simplifying assumption of vanishing double discontinuity made in this paper. Let us comment on this next step in some detail. 

The Mandelstam representation of the amplitude takes the form
\begin{equation}
  \label{eq:mandelstamRep1}
  \begin{aligned}
T(s,t) = c_0 + \int_{4 m^2}^\infty \frac{d s'}{\pi} \frac{\rho(s')}{s'-s}+(s\leftrightarrow t)+(s\leftrightarrow u) + \\ + \int_{4 m^2}^\infty \frac{d s' d t'}{\pi^2}
\frac{\rho(s',t')}{(s'-s)(t'-t)}  + (s,u) + (t,u) \ ,
\end{aligned}
\end{equation}
and the main novelty compared to the present work is the presence of the double discontinuity $\rho(s,t)$. The original Atkinson-Mandelstam problem then considers as an input $c_0$ and inelasticity which is encoded in $\eta_{\text{MP}}(s) \equiv 1 - |S_0(s)|^2$ and $\rho_{\text{MP}}(s,t)$, see \cite{Tourkine:2023xtu} for further details.

In~\cite{Tourkine:2023xtu}, we devised a fixed-point algorithm that solves this problem numerically for given $\left(c_0, \eta_{\text{MP}}, \rho_{\text{MP}}(s,t) \right)$ by iterations. Making this method practical, however, requires addressing three further issues: how do we improve the convergence range of the Atkinson-Mandelstam iterative algorithm? How do we effectively scan over the space of scattering amplitudes? How do we impose unitarity $|S_J(s)| \leq 1$ above the multi-particle thresholds?

We believe that neural optimizer offers a natural way to address all these problems. First, as we have seen in the present paper, the gradient-descent method combined with the NN parameterization of the amplitude allowed us to cover the full space of amplitudes. We expect this to continue in the full problem.
Second, to scan over the space of amplitudes we could use two NNs parametrizing the single spectral density and the double spectral density respectively. In particular, what was previously inputs of iterative method: $\eta_{\text{MP}}$ and $\rho_\text{MP}$, would be described by the NNs as well, and could be dynamically updated during learning process of solving unitarity equations.\footnote{A related interesting direction concerns further refinement of the bootstrap algorithm coming from (at least partial) implementation of multi-particle unitarity, see, e.g.,~\cite{Correia:2021etg}.} 
The same comment applies to the possibility of scanning over the space of low-energy Taylor coefficients of the amplitude, as explicitly demonstrated in the present paper. 
Finally, as we discussed around \eqref{eq:uniineq}, imposing unitarity inequalities above the multi-particle threshold is straightforward by adding extra terms to the loss function.

The main technical difficulty in implementing the full problem is the presence of the double integrals in the Mandelstam representation and in the Mandelstam unitarity equations, in particular. We must discretize these double integrals to reduce the problem to tensor multiplications and apply the gradient-descent neural optimizer. For this problem, machine learning will also be useful from another perspective, not used in this paper: most algorithms are optimized to run on GPUs, which will prove extremely useful for this more intensive numerical task.\footnote{The neural optimizers of the present paper can be run on GPUs, as documented in the ancillary files.} We hope to report on progress in this direction soon. 

From a physics perspective, the dispersive bootstrap methods developed in this paper have the potential to shed further light on the microscopic, possibly Lagrangian, origin of the amplitudes generated in the bootstrap studies. As a concrete example, in the full almond in Figure~\ref{fig:full_region}, the microscopic origin of the amplitudes that populate it beyond the weakly coupled $\phi^4$ region close to the origin is not well understood. Having better control over the structural properties of the amplitude, namely the support of the double-spectral density, clean separation of the truly multi-particle, and quasi-elastic physics could help us to connect the bootstrap results to known microscopic properties of physical amplitudes.

\bigskip

\noindent\textit{Acknowledgements.}
We thank Aurelien Dersy, Christopher Eckner, Andrea Guerrieri, Slava Rychkov, Matthew Schwartz, and participants of the ``2024 Mini-workshop on the S-matrix bootstrap'' at LAPTh in Annecy for useful discussions and comments. PT would like to thank Michael Kagan for discussions on machine learning related to this problem at an earlier stage.
We thank the Centre de Physique Théorique Grenoble Alpes for organizing a cross-physics AI workshop, during which we gathered ideas to improve our NN technology. This work has received funding from Agence Nationale de la Recherche (ANR), project ANR-22-CE31-0017. This project has
received funding from the European Research Council (ERC) under the European Union’s
Horizon 2020 research and innovation program (grant agreement number 949077). The drawings were made using the editor IPE~\cite{ipe}.


\appendix
\addtocontents{toc}{\protect\setcounter{tocdepth}{1}} 

\section{Dual search in the $(c_0,c_2)$ plane}
\label{sec:c0c2plane}
At the beginning of Section \ref{sec:Dual}, we described the extremization problem of a single Taylor coefficient $c_0$ of the amplitude. Here we explain how this procedure can be generalized to explore the $(c_0,c_2)$ plane. Let us modify the primal Lagrangian~\eqref{eq:primal_lag} to be
\begin{align}
    \mathcal{O} &= \alpha_0 c_0 + \alpha_2 c_2 + \kappa_0 \left[c_0 {-} \int^\infty_{4m^2} \!\!\! dv \, \frac{3}{\pi} \frac{ {n_0} \, \im f_0(v)}{(v{-}\tfrac{4m^2}{3})} \right] + \kappa_2 \left[c_2 {-} \int^\infty_{4m^2} \!\!\! dv \, \frac{1}{\pi} \frac{ n_0 \, \im f_0(v)}{(v{-}\tfrac{4m^2}{3})^3} \right] \nonumber \\
    &+ \int^{\mu^2}_{4m^2} \!\!\! ds \, w_0(s) \, \mathcal{A}_0(s) + \int^\infty_{4m^2} \!\!\! dv \, \lambda_0(v) \, n_0^2 \, \det \mathcal{U}_0(v) \,,
    \label{eq:primal_lag_twod}
\end{align}
and $\kappa_2$ is an additional dual variable. The variables $(\alpha_0,\alpha_2)$ are constants chosen so that the dual problem finds the maximal value of $\alpha_0 c_0+\alpha_2 c_2$, allowing us to shoot at an angle in the $(c_0,c_2)$ plane.
The equations of motion \eqref{eq:Kappa_eqofmot} for the primal variables $\{c_0,c_2\}$ take the following new form:
\begin{equation}
    {\alpha_2} + \kappa_2 = 0 \qquad \text{and} \qquad {\alpha_0} + \kappa_0 - \int_{4m^2}^{\mu^2} \!\!\! ds \, \frac{w_0(s)}{{ 16\pi }} = 0 \, .
\label{eq:Kappa_eqofmot_twod}
\end{equation}
Analogously, the auxiliary function~\eqref{eq:Mubar} becomes
\begin{align}
    {\overline \mu}_0(v) &\equiv -\frac{1}{\pi} \left(\frac{3\kappa_0}{(v{-}\tfrac{4m^2}{3})} + \frac{\kappa_2}{(v{-}\tfrac{4m^2}{3})^3}\right) - \text{P.V.} \!\! \int^{\mu^2}_{4m^2} \!\!\! ds \, w_0(s) \, k_{0,0}(s,v) \, .
    \label{eq:Mubar_twod}
\end{align}

Above setup is useful to extremize in a particular direction in the $(c_0,c_2)$ plane fixed by the values of $(\alpha_0, \alpha_2)$, which would allow us to scan the allowed region radially. 

Alternatively, we can do a vertical scan i.e. we fix $c_0$ equal to various values of $c_0^\text{fixed} \in [0,2.66]$ and maximize/minimize the value of $c_2$. This is the procedure we choose in this work, and we connect the vertical slices obtained for each $c_0^\text{fixed}$ to draw the Figure~\ref{fig:results}.

This procedure requires an extra linear constraint $c_0 - c_0^\text{fixed} = 0$ added to the setup. We can implement it by
\begin{equation}
    \mathcal{O} = \tilde{\alpha}_0 (c_0-c_0^\text{fixed}) + c_2 + \text{Lagrange multipliers of }\eqref{eq:primal_lag_twod} \quad ,
\end{equation}
and promoting $\tilde{\alpha}_0$ to be a yet another free dual variable to be optimized in.

Note that this setup can possibly shift $\overline{\mathcal{D}}$ to take negative values, due to the new term $-\tilde{\alpha}_0 c_0^\text{fixed}$. In Section~\ref{sec:Dual_NN}, we explain how we carefully deal with this case, while choosing a positive definite loss function in the NN.

\section{Neural network details}

In this appendix, we provide details about the grid we used for the energy variable $x$ and the numerical integration method we used to perform the dispersive integral.

\subsection{Grid}
\label{sec:grid}
After applying the map $x(s)= 4m^2/s$, we discretize the interval $[0,1]$ with $N+1$ points
\begin{equation}
    0=x_0<x_1<\cdots<x_{N-1}<x_N=1.
\end{equation}
The grid used in Section~\ref{sec:Primal} for the primal problem consists of: $N_0=450$ points linearly spaced between 0 and 1; $N_1=300$ points logarithmically-spaced between $10^{-100}$ and $1$; $N_2=60$ points logarithmically-spaced between $1-10^{-20}$ and $0$, such that logarithmic densities are equal. This grid contains a total of 810 points.

For the dual strategy (Section~\ref{sec:Dual_NN}) since we incorporated the sum rule~\eqref{eq:c0-sumrule} as a constraint (with $\kappa_0 \neq 0$) which relates the UV to the IR (as discussed in Section~\ref{sec:uv-ir-sens}), it was not necessary to include high-energy points to achieve good convergence. 

We used the same number of logarithmically spaced grid points, $N_1 = 300$, near both the Regge and threshold regions, with cutoffs at $10^{-8}$ and $1-10^{-8}$, respectively. Additionally, $N_0=300$ linearly spaced points were used, resulting in a grid with a total of 900 points. 

\subsection{Numerical integration for the neural network}
\label{sec:integration_matrix}
Let us explain how the integrations in Section~\ref{sec:Dual_NN} and Section~\ref{sec:Primal_NN} are computed during the training of the neural network. As explained in the text, the backpropagation algorithm must compute the gradient of the loss~(\ref{eq:backpropagation}). Therefore, only operations that are differentiable and compatible with PyTorch's automatic differentiation can be applied to the network's output when constructing the loss. These operations include smooth, elementary functions and standard tensor operations, as they ensure the computation of gradients required for optimization. We transformed every integral appearing in the loss into matrix multiplication to enable the computation of the gradient. As an example, let us consider the dispersion relations \eqref{eq:projected_disp}, recalled here
\begin{equation}
    \re f_0(s) = \dfrac{c_0}{16\pi}+\mathrm{P.V.}\!\!\int_{4m^2}^{\infty}\!\!\!\mathrm{d}v\, {n_0} \, k_{0,0}(s,v)\,\im f_0(v),\quad s\in[4m^2,\infty)\,.\label{eq:disp_appendix}
\end{equation}
The same method is applied to compute integrals encountered in the dual strategy. For the purposes of this discussion, the principal value term will be omitted in subsequent equations as it is not relevant to the explanation.

We first rewrite Eq.~\eqref{eq:disp_appendix} in terms of the variables $x=4m^2/s$ and $y=4m^2/v$ 
\begin{equation}
    \re f_0(x) = \dfrac{c_0}{16\pi}+\int_{0}^{1}\!\!\!\mathrm{d}y\,{n_0} \, \tilde{k}_{0,0}(x,y)\,\im f_0(y),
\end{equation}
where the kernel $\tilde{k}_{0,0}(x,y)$ is defined by
\begin{equation}
    \tilde k_{0,0}(x,y) = \dfrac{4m^2}{y^2}\, k_{0,0}\left(\frac{4m^2}x,\frac{4m^2}y\right)\,.
\end{equation}

Next, we use the grid defined in Appendix~\ref{sec:grid} to discretize both the $y$-integration and the $x$-variable:
\begin{equation}
    \re f_0(x_n) = \dfrac{c_0}{16\pi}+\sum_{k=1}^{N}\int_{x_{k-1}}^{x_k}\!\!\!\mathrm{d}y\, {n_0} \, k_{0,0}(x_n,y)
    g(y).
    \label{eq:ReF_x}
\end{equation}
where, on each segment $[x_{k-1},x_k]$, the function $g(y)$ is a linear interpolation of $\im f_0$, defined by:
\begin{equation}
    \forall y\in\left[x_{k-1},\,x_k\right],\quad
    g(y)=\im f_0(x_{k-1})+\left(\im f_0(x_k)-\im f_0(x_{k-1})\right)\dfrac{y-x_{k-1}}{x_k-x_{k-1}}.\label{eq:linear_interpolation}
\end{equation}
\begin{subequations}
By combining eqs~\eqref{eq:ReF_x}~and~\eqref{eq:linear_interpolation}, we obtain
\begin{equation}
    \re f_0(x_n) = \dfrac{c_0}{16\pi}+\sum_{k=1}^{N-1}\im f_0(x_k)\,{n_0}\!\left(\int_{x_{k-1}}^{x_k}\!\!\!\!\!\!\!\mathrm{d}y\, k_{0,0}(x_n,y)\dfrac{y-x_{k-1}}{x_k-x_{k-1}}+\int_{x_{k}}^{x_{k+1}}\!\!\!\!\!\!\!\!\mathrm{d}y\,k_{0,0}(x_n,y)\dfrac{x_{k+1}-y}{x_{k+1}-x_k}\right),\label{eq:disp_discretized}
\end{equation}
where we assumed that $\im f_0(x_0)=\im f_0(x_N)=0$.
Eq.~\eqref{eq:disp_discretized} can finally be expressed in a matrix form as
\begin{equation}
    \re \vec{f_0} = \dfrac{c_0}{16\pi}+M\cdot\im \vec{f_0}\,,
   \label{eq:disp_matrix}
\end{equation}
where the matrix elements $M_{nk}$ are given by
\begin{equation}
    M_{nk} = {n_0}\!\!\int_{x_{k-1}}^{x_k}\!\!\!\!\!\!\!\mathrm{d}y\, k_{0,0}(x_n,y)\dfrac{y-x_{k-1}}{x_k-x_{k-1}}+{n_0}\!\!\int_{x_{k}}^{x_{k+1}}\!\!\!\!\!\!\!\!\mathrm{d}y\,k_{0,0}(x_n,y)\dfrac{x_{k+1}-y}{x_{k+1}-x_k}.
\end{equation}
\end{subequations}

When working with the singular threshold, we do not have anymore that $\im f_0(x_N)=0$ (recall $\im f_0(s)\sim \sqrt{s-4m^2}^{-1}$). To accommodate this, we decompose $\im f_0(x)$ into a regular and a singular part, to arrive at the ansatz~\eqref{eq:dressed_NN_sing}:
\begin{subequations}
    \begin{equation}
    \im f_0(x)=\im f_\mathrm{sing}(x)+\im f_\mathrm{reg}(x),\label{eq:reg+sing}
\end{equation}where
\begin{equation}
    \im f_\mathrm{sing}(s) = 2\dfrac{\mathcal{R}(s)}{\varphi(s)},
\end{equation}and
\begin{equation}
    \im f_\mathrm{reg}(s) = 2\dfrac{\mathcal{R}(s)}{\varphi(s)}\mathrm{CELU\left(\varphi(s)^2\mathrm{NN}_\sigma(\frac{4m^2}{s})\right)}.
\end{equation}
Crucially, only $\im f_\mathrm{reg}$ depends on the neural network output and satisfies $\im f_\mathrm{reg}(x_N)=0$.
\end{subequations}
The integral in~\eqref{eq:ReF_x} can be computed once and for all for $\im f_\mathrm{sing}$, yielding $\re f_\mathrm{sing}$.
To~compute $\re f_\mathrm{reg}(x)$ from $\im f_\mathrm{reg}(x)$ we  use \eqref{eq:disp_discretized}, since \begin{equation}
    \im f_\mathrm{reg}(x_0)=\im f_\mathrm{reg}(x_N)=0.
\end{equation}

\section{Semi-definite programming details}

In this appendix, we explain the details of our implementation of the primal and dual bootstrap using the numerical tools of semi-definite programming.

\subsection{Primal}
\label{sdpb_details_primal}

Let us explain how we implemented the primal bootstrap discussed in Section~\ref{sec:rho_bootstrap}. 

We use the $\rho$-wavelets in \eqref{eq:primal_ansatz} as our basis to span the space of amplitudes. They are convenient one-to-one maps between the unit disc and the physical sheet. Let us give first the inverse of \eqref{eq:rho-map} as a function of wavelet center $\sigma$:
\begin{equation}
    (\rho_\sigma)^{-1}: \rho \mapsto \frac{8 \left(\rho ^2+1\right)-(\rho -1)^2 \sigma }{(\rho +1)^2} \, .
\end{equation}
We impose the unitarity conditions by sampling $\mathcal{U}_0(s)$ on \emph{a unitarity grid} of points $\{s_k\}$  with $k \in \{ 1 ,\dots, n_\text{pts} \}$, defined through the image of a Chebyshev grid $\{\phi_k\}$ on the boundary of the unit disc centered at $\sigma=20/3$, as follows
\begin{equation}
    s_k = (\rho_{20/3})^{-1}\left[\exp(\phi_k)\right] \quad , \quad \phi_k = \frac{\pi}{2}\left(1+\cos\left[\tfrac{k \pi}{n_\text{pts}+1}\right]\right)
\end{equation}
We chose the value of $n_\text{pts}=300$, and we also checked that increasing the grid density does not change the results.

Then, we chose a certain subset of the unitarity grid to center the basis functions $\rho_\sigma$ with free coefficients $\alpha_\sigma$. 
We defined this subset $\Sigma_N$ as follows
\begin{equation}
    \Sigma_N=\bigcup_{n=1}^N \Sigma_n \quad , \quad \Sigma_n = \left\{ (\rho_{20/3})^{-1}\left[\exp(\phi_k)\right]
    \, | \, k=1 \dots n \right\}
\end{equation}
We ran our program until we reach convergence around $N=12$.

\subsection{Dual}
\label{sdpb_details_dual}
Here we explain how we implemented the dual bootstrap problem discussed in Section \ref{sec:dual_sdpb}. For brevity, we set $m^2=1$ in this appendix.

We use Chebyshev polynomials $T_n(x)$ to span the space of dual functions $\{ w_0, \mathcal{X}^\text{IR}_0, \mathcal{X}^\text{UV}_0 \}$. To accomplish that, we first map the interval $x \in [-1,1]$ onto $v^\text{IR} \in [4,\mu^2]$ and $v^\text{UV} \in [\mu^2,\infty)$ as follows
\begin{equation}
\begin{aligned}
v^\text{IR}(x) = \frac{1}{2} \left( \, (\mu^2-4)x + \mu^2 + 4 \, \right) \quad , \quad v^\text{UV}(x) = \frac{20 + (20-2\mu^2) \sin \left[ \tfrac{\pi}{2} x \right]}{\cos \left[ \tfrac{\pi}{2} (x+1) \right]+1} \, .
\end{aligned}
\end{equation}
Then, on the compact support of $x$, we use
\begin{equation}
\begin{aligned}
w_0(x) = \sum^{N_w}_{n=0} a_{n} \, T_n(x) \quad , \quad \mathcal{X}^{\text{IR}}_0(x) = \sum^{N_\text{IR}}_{n=0} c_{n} \, T_n(x) \quad , \quad \mathcal{X}^{\text{UV}}_0(x) = \sum^{N_\text{UV}}_{n=0} d_{n} \, T_n(x)
\label{dual_ansatzes_o1}
\end{aligned}
\end{equation}
such that $dx \, w_0(x) = dv \, w_0(v)$. Since the Jacobian $dv^\text{UV}/dx$ has a zero at $x=-1$, in order to stop $x$ integrands from diverging, we require
\begin{equation}
\mathcal{X}^{\text{UV}}_0(-1) = \sum^{N_\text{UV}}_{n=0} d_{n} \, T_n(-1) = 0 \, . \label{continuity_o1}
\end{equation}
In all our runs, we sample these functions on a Chebyshev grid $\{x_k\}$ defined as,
\begin{equation}
x_k = \cos\left[\tfrac{k \pi}{n_\text{pts}+1}\right] \, \text{ where } \, k \in \{1,\dots, n_\text{pts}\},
\label{cheby_grid}
\end{equation}
for both IR and UV sections. We ran with both $n_\text{pts}=199$ and $999$ to make sure that the solutions for dual functions \eqref{dual_ansatzes_o1} stabilize with respect to a denser sampling.  Let us also remark that, a sufficiently dense grid is enough to enforce the dual conditions \eqref{eq:SD_conds} for all $v \in [4,\infty)$, which we cross-check by evaluating \eqref{dual_ansatzes_o1} with fixed coefficients $\{a_n,c_n,d_n\}$ on new points outside of the grid. The dual bounds then have no systematic error due to discretization, hence they keep being rigorous.

All in all, the discretization of the dual problem \eqref{eq:min_dlin} leads us to
\begin{equation}
\begin{aligned}
\min \quad \overline{\mathcal{D}}^\text{lin} \qquad &\text{over} \quad \{ a_{n},c_{n},d_{n} \} \quad \text{subject to} \quad \{ \eqref{eq:SD_conds}|_{x \in \{x_k\}} \, , \, \eqref{continuity_o1} \}
\end{aligned}
\end{equation}
We always set
\begin{equation}
N_w = N_\text{IR} \equiv N_\text{max} \qquad , \qquad N_\text{UV} = 0 \, .
\end{equation}
and run our program until we reach convergence around 
$N_\text{max}=96$.

\bibliographystyle{JHEP}
\bibliography{bib-papers} 

\end{document}